\documentclass[12pt]{article}
\usepackage{amssymb,amsmath,epsfig}
\allowdisplaybreaks

\begin{document}

\title{\bf Influence of Charge on Extended Decoupled Anisotropic
Solutions in
$f(\mathcal{R},\mathcal{T},\mathcal{R}_{\lambda\xi}\mathcal{T}^{\lambda\xi})$
Gravity}
\author{M Sharif$^{1}$ \thanks{Corresponding author, E-mail: msharif.math@pu.edu.pk} and T Naseer$^{1}$\\
$^{1}$ Department of Mathematics, University of the Punjab,\\
Quaid-i-Azam Campus, Lahore-54590, Pakistan.}

\date{}
\maketitle

\begin{abstract}
In this paper, we consider static self-gravitating spherical
spacetime and determine various anisotropic solutions through the
extended gravitational decoupling technique in
$f(\mathcal{R},\mathcal{T},\mathcal{R}_{\lambda\xi}\mathcal{T}^{\lambda\xi})$
gravity to analyze the influence of electromagnetic field on them.
We construct two different sets of modified field equations by
employing the transformations on both radial as well as temporal
metric potentials. The first set symbolizes the isotropic fluid
distribution, thus we take Krori-Barua solution to deal with it. The
indefinite second sector comprises the influence of anisotropy. In
this regard, we apply some constraints to determine unknowns.
Further, we observe the impact of charge as well as decoupling
parameter $\zeta$ on the developed physical variables (such as
energy density,~radial and tangential pressures) and anisotropy. We
also analyze other physical features of the compact geometry like
mass, compactness and redshift along with the energy conditions.
Eventually, we find that our both solutions show less stable
behavior for higher values of charge near the boundary in this
gravity.
\end{abstract}

\section{Introduction}

The composition of our universe is well-structured yet inscrutable,
comprising of heavily geometrical objects like stars, galaxies and
other unfathomable ingredients. The study of physical features of
such massive structures plays an important role to figure out cosmic
evolution. Einstein developed first General Relativity (GR) which
allows scientists to get better understanding of both cosmological
as well as astrophysical phenomena. Several cosmological experiments
have been performed on the distant galaxies which indicate that our
universe is in the state of accelerating expansion. This expansion
is thought to be executed by the dark energy which is repulsive in
nature. The modifications to GR have therefore been identified
highly significant to reveal mysterious characteristics of our
cosmos. The simplest modification to GR is obtained by replacing the
Ricci scalar $\mathcal{R}$ with its generic function in an
Einstein-Hilbert action, named as $f(\mathcal{R})$ theory. Numerous
research \cite{2}-\cite{2c} has been done in this theory to analyze
the physical feasibility of compact structures through different
techniques. The Lan\'{e}-Emden equation in $f(\mathcal{R})$ theory
has been employed by Capozziello \emph{et al.} \cite{8} to discuss
the stability of various mathematical models. Many authors
\cite{9}-\cite{9g} studied different astronomical objects and
examined their formation as well as stable configuration.

Initially, the concept of matter-geometry coupling was introduced by
Bertolami \emph{et al.} \cite{10}. They studied the effects of
coupling in $f(\mathcal{R})$ gravity by taking the Lagrangian as a
function of $\mathcal{R}$ and $\mathcal{L}_{m}$. This kind of
interaction in modified theories has prompted many researchers to
put their concentration on accelerating nature of cosmic expansion.
Many other modified theories have been developed in the last decade
to investigate the role of arbitrary matter-geometry interaction on
massive structures, one of them is the $f(\mathcal{R},\mathcal{T})$
theory introduced by Harko \emph{et al.} \cite{20} in which
$\mathcal{T}$ expresses trace of the stress-energy tensor. Such
interaction provides the non-conserved energy-momentum tensor which
may lead to the accelerated expansion of universe. Later, another
theory involving more complex functional was presented by Haghani
\emph{et al.} \cite{22}, named as
$f(\mathcal{R},\mathcal{T},\mathcal{Q})$ gravity, where
$\mathcal{Q}\equiv
\mathcal{R}_{\lambda\xi}\mathcal{T}^{\lambda\xi}$. The conserved
equations of motion have also been found through Lagrange multiplier
approach in this theory. In this scenario, Sharif and Zubair assumed
some mathematical models in FRW spacetime and figured out the black
hole laws of thermodynamics \cite{22a} as well as energy bounds
\cite{22b}.

The development of this modified gravity was premised on the
insertion of the factor $\mathcal{Q}$ which ensures the presence of
strong non-minimal matter-geometry coupling in self-gravitating
systems. The modification in the Einstein-Hilbert action may help in
explaining the role of dark energy and dark matter, without
resorting to exotic fluid distribution. Some other extensions to GR
like $f(\mathcal{R},\mathcal{L}_m)$ and $f(\mathcal{R},\mathcal{T})$
gravitational theories also comprise the matter Lagrangian involving
such arbitrary interaction but we cannot consider their functionals
as the most generalized form that provide proper understanding to
the influence of coupling on self-gravitating objects in some
scenarios. It must be noted here that the factor
$\mathcal{R}_{\lambda\xi}\mathcal{T}^{\lambda\xi}$ could explain the
impact of non-minimal interaction in the situation where
$f(\mathcal{R},\mathcal{T})$ theory breaks down to achieve such
results. In particular, $f(\mathcal{R},\mathcal{T})$ theory does not
provide coupling effects on the gravitational model for the case
when trace-free energy-momentum tensor (i.e., $\mathcal{T}=0$) is
considered, however, this phenomenon can be explained by
$f(\mathcal{R},\mathcal{T},\mathcal{Q})$ gravity. Due to the
non-conservation of energy-momentum tensor in this theory, an
additional force is present due to which the motion of test
particles in geodesic path comes to an end. This force also helps to
elucidate the galactic rotation curves.

Haghani \emph{et al.} \cite{22} considered three different models
such as
$\mathcal{R}+\varrho\mathcal{Q},~\mathcal{R}(1+\varrho\mathcal{Q})$
and $\mathcal{R}+\gamma\sqrt{\mid
\mathcal{T}\mid}+\varrho\mathcal{Q}$ in this framework and discussed
their cosmological applications, where $\varrho$ and $\gamma$ are
arbitrary coupling constants. The dynamics and cosmic evolution has
been explored with respect to these models with and without
considering the energy conservation. Odintsov and S\'{a}ez-G\'{o}mez
\cite{23} studied various models in
$f(\mathcal{R},\mathcal{T},\mathcal{Q})$ theory and solved their
corresponding gravitational equations through numerical methods.
They also highlighted some serious issues linked with the matter
instability. Ayuso \emph{et al.} \cite{24} obtained the conditions
for different compact objects to be stable in this theory by
considering some suitable scalar and vector fields. They concluded
that the existence of matter instability is necessary for the case
of vector field. For FRW geometry, Baffou \emph{et al.} \cite{25}
incorporated the perturbation functions to calculate the solution of
modified field equations and checked their viability. Yousaf
\emph{et al.} \cite{26}-\cite{26e} studied the evolution of
charged/uncharged spherical as well as cylindrical geometries with
the help of effective structure scalars.

The existence of electromagnetic field in celestial structures helps
to understand their expansion and stability in a better way.
Numerous investigations have been performed in GR as well as
modified theories to analyze the role of charge on different
physical properties of celestial objects. The Einstein field
equations coupled with the charge are known as Einstein-Maxwell
equations whose solution has been found by Das \emph{et al.}
\cite{27} through matching the static interior spacetime with
exterior Reissner-Nordstr\"{o}m. Sunzu \emph{et al.} \cite{27a}
examined the quark stars influenced from charge by utilizing the
mass-radius relation. Murad \cite{28} analyzed the charged strange
stars having anisotropic configuration and studied its physical
characteristics. Many authors \cite{28a}-\cite{28h} analyzed
different stellar structures and observed their more stable behavior
due to the presence of charge.

The nature of field equations corresponding to the self-gravitating
body is highly non-linear, thus it is much difficult task to obtain
their exact solution. There has been various techniques to solve
these equations in the literature corresponding to the celestial
objects coupled with isotropic as well as anisotropic
configurations. Sharif and Waseem \cite{28i,28j} investigated the
viability and stability of several compact objects in
curvature-matter coupled gravity by using Krori-Barua ansatz. They
found that solutions for anisotropic matter distribution are stable,
while isotropic configured stars are shown to be unstable near the
core. Maurya \emph{et al.} \cite{28k} considered the minimal
coupling model as
$f(\mathcal{R},\mathcal{T})=\mathcal{R}+2\chi\mathcal{T}$ ($\chi$ is
served as the coupling parameter) and examined the physical
viability of anisotropic structures through embedding class one
condition along with MIT bag model. Several compact anisotropic
configurations have been discussed by Shamir and Fayyaz \cite{28l}
in $f(\mathcal{R})$ theory by taking different models. The newly
developed method, namely minimal geometric deformation (MGD) by
means of gravitational decoupling has been found to be significant
to develop physically feasible solutions in the field of cosmology
and astrophysics. This technique was initially introduced by Ovalle
\cite{29} in the braneworld scenario to develop exact solutions for
spherical interstellar structures. The formulation of analytical
solutions for isotropic geometry has been done by Ovalle and Linares
\cite{30} in the context of braneworld. They also shown their
compatibility with the Tolman IV solution. Casadio \emph{et al.}
\cite{31} calculated a new solution for spherical exterior geometry
which shows singular behavior at Schwarzschild radius.

Ovalle \cite{32} utilized the decoupling scheme to obtain
anisotropic exact solution for a compact sphere. Ovalle \emph{et
al.} \cite{33} considered the isotropic solution and extended it to
various anisotropic solutions whose feasibility has been analyzed
graphically. Gabbanelli \emph{et al.} \cite{34} formulated
physically acceptable anisotropic solution by assuming
Duragpal-Fuloria isotropic ansatz. Estrada and Tello-Ortiz
\cite{34a} considered Heintzmann solution for isotropic distribution
and found different physically viable anisotropic solutions. Sharif
and Sadiq \cite{35} employed Krori-Barua anstaz and determined two
anisotropic solutions through this method. They also analyzed the
influence of decoupling parameter and charge on physical parameters
as well as energy bounds. Sharif and his collaborators
\cite{36}-\cite{36c} generalized different anisotropic solutions to
modified theories such as $f(\mathcal{G})$ and $f(\mathcal{R})$,
where $\mathcal{G}$ represents the Gauss-Bonnet invariant. Singh
\emph{et al.} \cite{37} utilized class one condition to obtain
various anisotropic solutions with the help of same strategy. Hensh
and Stuchl\'{i}k \cite{37a} constructed different feasible
anisotropic solutions by employing a suitable deformation function
on the field equations along with Tolman VII solution.

Although the MGD technique (in which the radial metric component is
transformed only, while the temporal component remains preserved) is
an immensely valuable approach to find exact feasible solutions of
the complicated field equations. This is nevertheless possible only
if energy is not exchanged from one source to another. Ovalle
\cite{38} addressed this issue by proposing a novel strategy, named
as extended gravitational decoupling (EGD). This technique
transforms both (radial and temporal) metric potentials and also
works for any choice of fluid distribution for all spacetime
regions. Through this scheme, Contreras and Bargue\~{n}o \cite{39}
presumed vacuum BTZ solution corresponding to the $2+1$-dimensional
geometry and found its extension for the charged case. Sharif and
Ama-Tul-Mughani have employed this method along with the isotropic
Tolman IV \cite{40} as well as charged Krori-Barua \cite{41} anstaz
and constructed various anisotropic solutions. Sharif and Saba
\cite{41a} calculated anisotropic spherical solutions by considering
Tolman IV in $f(\mathcal{G})$ gravity. We have recently checked the
physical feasibility of two charged/uncharged anisotropic solutions obtained
through MGD approach in $f(\mathcal{R},\mathcal{T},\mathcal{Q})$
theory \cite{41b,41c}. Sharif and Majid \cite{41d}-\cite{41f} formulated
various cosmological solutions by extending the isotropic
Krori-Barua and Tolman IV solutions with the help of MGD as well as
EGD techniques in the framework of Brans-Dicke gravity.

This paper examines the effects of charge on various anisotropic
solutions constructed through EGD technique in
$f(\mathcal{R},\mathcal{T},\mathcal{Q})$ scenario. The paper has the
following structure. In the next section, we shall present some
fundamental terminologies of this modified theory. Section 3
discusses the EGD technique which helps to split the field equations
into two independent sets by deforming the radial as well as
temporal metric components. We assume the Krori-Barua solution in
section 4 and employ some additional constraints to formulate two
anisotropic solutions. We also discuss their viability and stability
by means of graphs. Lastly, we sum up our results in section 5.

\section{The $f(\mathcal{R},\mathcal{T},\mathcal{Q})$ Gravity}

The modified Einstein-Hilbert action (with $\kappa=8\pi$) influenced
by an additional source is given as \cite{23}
\begin{equation}\label{g1}
S_{f(\mathcal{R},\mathcal{T},\mathcal{R}_{\lambda\xi}\mathcal{T}^{\lambda\xi})}=\int
\left[\frac{f(\mathcal{R},\mathcal{T},\mathcal{R}_{\lambda\xi}\mathcal{T}^{\lambda\xi})}{16\pi}
+\mathcal{L}_{m}+\mathcal{L}_{\mathcal{E}}+\zeta\mathcal{L}_{\Upsilon}\right]\sqrt{-g}d^{4}x,
\end{equation}
where $\mathcal{L}_{m},~\mathcal{L}_{\mathcal{E}}$ and
$\mathcal{L}_{\Upsilon}$ indicate the Lagrangian densities of matter
configuration, electromagnetic field and new gravitationally coupled
source, respectively. In this case, we take $\mathcal{L}_{m}=-\mu$,
$\mu$ is the energy density of the fluid. The field equations
analogous to the action \eqref{g1} have the form as
\begin{equation}\label{g2}
\mathcal{G}_{\lambda\xi}=8\pi \mathcal{T}_{\lambda\xi}^{(tot)}.
\end{equation}
The geometry of the celestial structure is expressed by an Einstein
tensor $\mathcal{G}_{\lambda\xi}$ whereas
$\mathcal{T}_{\lambda\xi}^{(tot)}$ is the energy-momentum tensor
including all sources. We further write it as
\begin{equation}\label{g3}
\mathcal{T}_{\lambda\xi}^{(tot)}=\mathcal{T}_{\lambda\xi}^{(eff)}+\zeta
\Upsilon_{\lambda\xi}=-\frac{1}{\mathcal{L}_{m}f_{\mathcal{Q}}-f_{\mathcal{R}}}\left(\mathcal{T}_{\lambda\xi}^{(m)}
+\mathcal{E}_{\lambda\xi}\right)+\mathcal{T}_{\lambda\xi}^{(\mathcal{D})}+\zeta\Upsilon_{\lambda\xi}.
\end{equation}
The presence of new source $\Upsilon_{\lambda\xi}$ via certain
scalar, vector or tensor field generates anisotropy in the
self-gravitating body. The decoupling parameter $\zeta$ measures how
much that source affects the geometry. In addition, the quantity
$\mathcal{T}_{\lambda\xi}^{(eff)}$ can be viewed as the
stress-energy tensor in $f(\mathcal{R},\mathcal{T},\mathcal{Q})$
gravity which comprises the usual physical variables as well as
modified correction terms. In this scenario, the modified sector
$\mathcal{T}_{\lambda\xi}^{(D)}$ takes the form
\begin{eqnarray}
\nonumber
\mathcal{T}_{\lambda\xi}^{(\mathcal{D})}&=&-\frac{1}{8\pi\big(\mathcal{L}_{m}f_{\mathcal{Q}}-f_{\mathcal{R}}\big)}
\left[\left(f_{\mathcal{T}}+\frac{1}{2}\mathcal{R}f_{\mathcal{Q}}\right)\mathcal{T}_{\lambda\xi}^{(m)}
+\left\{\frac{\mathcal{R}}{2}(\frac{f}{\mathcal{R}}-f_{\mathcal{R}})-\mathcal{L}_{m}f_{\mathcal{T}}\right.\right.\\\nonumber
&-&\left.\frac{1}{2}\nabla_{\rho}\nabla_{\eta}(f_{\mathcal{Q}}\mathcal{T}^{\rho\eta})\right\}g_{\lambda\xi}
-\frac{1}{2}\Box(f_{\mathcal{Q}}\mathcal{T}_{\lambda\xi})-(g_{\lambda\xi}\Box-
\nabla_{\lambda}\nabla_{\xi})f_{\mathcal{R}}\\\label{g4}
&-&2f_{\mathcal{Q}}\mathcal{R}_{\rho(\lambda}\mathcal{T}_{\xi)}^{\rho}+\nabla_{\rho}\nabla_{(\lambda}[\mathcal{T}_{\xi)}^{\rho}f_{\mathcal{Q}}]
+2(f_{\mathcal{Q}}\mathcal{R}^{\rho\eta}+\left.f_{\mathcal{T}}g^{\rho\eta})\frac{\partial^2\mathcal{L}_{m}}
{\partial g^{\lambda\xi}\partial g^{\rho\eta}}\right],
\end{eqnarray}
where $f_{\mathcal{R}},~f_{\mathcal{T}}$ and $f_{\mathcal{Q}}$ show
the derivatives of functional
$f(\mathcal{R},\mathcal{T},\mathcal{Q})$ with respect to their
subscripts. Moreover, $\nabla_\xi$ indicates the covariant
derivative and $\Box\equiv g^{\lambda\xi}\nabla_\lambda\nabla_\xi$.

The stress-energy tensor for perfect matter is given as
\begin{equation}\label{g5}
\mathcal{T}_{\lambda\xi}^{(m)}=(\mu+p) \mathcal{K}_{\lambda}
\mathcal{K}_{\xi}+pg_{\lambda\xi},
\end{equation}
where $p$ indicates the isotropic pressure and
$\mathcal{K}_{\lambda}$ is the four-velocity. The following equation
provides the trace of $f(\mathcal{R},\mathcal{T},\mathcal{Q})$ field
equations as
\begin{align}\nonumber
&3\nabla^{\xi}\nabla_{\xi}
f_\mathcal{R}+\mathcal{R}\left(f_\mathcal{R}-\frac{\mathcal{T}}{2}f_\mathcal{Q}\right)-\mathcal{T}(f_\mathcal{T}+1)+\frac{1}{2}
\nabla^{\xi}\nabla_{\xi}(f_\mathcal{Q}\mathcal{T})+\nabla_\lambda\nabla_\xi(f_\mathcal{Q}\mathcal{T}^{\lambda\xi})\\\nonumber
&-2f+(\mathcal{R}f_\mathcal{Q}+4f_\mathcal{T})\mathcal{L}_m+2\mathcal{R}_{\lambda\xi}\mathcal{T}^{\lambda\xi}f_\mathcal{Q}
-2g^{\rho\eta} \frac{\partial^2\mathcal{L}_m}{\partial
g^{\rho\eta}\partial
g^{\lambda\xi}}\left(f_\mathcal{T}g^{\lambda\xi}+f_\mathcal{Q}R^{\lambda\xi}\right)=0.
\end{align}
The assumption $\mathcal{Q}=0$ in the above equation vanishes strong
matter-geometry interaction in a self-gravitating object and
provides $f(\mathcal{R},\mathcal{T})$ theory, whereas one can
retrieve $f(\mathcal{R})$ theory by considering vacuum scenario. The
electromagnetic energy-momentum tensor takes the form
\begin{equation*}
\mathcal{E}_{\lambda\xi}=-\frac{1}{4\pi}\left[\frac{1}{4}g_{\lambda\xi}\mathcal{F}^{\rho\eta}\mathcal{F}_{\rho\eta}
-\mathcal{F}^{\eta}_{\lambda}\mathcal{F}_{\xi\eta}\right],
\end{equation*}
where the Maxwell field tensor is defined as
$\mathcal{F}_{\lambda\xi}=\omega_{\xi;\lambda}-\omega_{\lambda;\xi}$,
in which $\omega_{\xi}=\omega(r)\delta_{\xi}^{0}$ is the four
potential. This tensor must satisfy the following equations
\begin{equation*}
\mathcal{F}^{\lambda\xi}_{;\xi}=4\pi \mathcal{J}^{\lambda}, \quad
\mathcal{F}_{[\lambda\xi;\eta]}=0.
\end{equation*}
The current density $\mathcal{J}^{\lambda}$ can be expressed as
$\mathcal{J}^{\lambda}=\varrho \mathcal{K}^{\lambda}$, where
$\varrho$ is the charge density.

We consider a geometry which is separated into two regions, namely
interior and exterior at the hypersurface $\Sigma$. The static
spherically symmetric astrophysical structure is expressed by the
metric as follows
\begin{equation}\label{g6}
ds^2=-e^{\chi} dt^2+e^{\beta} dr^2+r^2d\theta^2+r^2\sin^2\theta
d\varphi^2,
\end{equation}
where $\chi=\chi(r)$ and $\beta=\beta(r)$. The above metric produces
the Maxwell field equations as
\begin{equation}
\omega''+\frac{1}{2r}\big[4-r(\chi'+\beta')\big]\omega'=4\pi\varrho
e^{\frac{\chi}{2}+\beta},
\end{equation}
which after integration produces
\begin{equation}
\omega'=\frac{q}{r^2}e^{\frac{\chi+\beta}{2}},
\end{equation}
where $q$ shows the charge in the interior geometry \eqref{g6}.
Here, $'=\frac{\partial}{\partial r}$. The four-velocity involves
single non-zero component due to the consideration of comoving
framework. Thus, it has the form
\begin{equation}\label{g7}
\mathcal{K}^\xi=(e^{\frac{-\chi}{2}},0,0,0).
\end{equation}
We establish the field equations in
$f(\mathcal{R},\mathcal{T},\mathcal{Q})$ theory corresponding to
sphere \eqref{g6} as
\begin{align}\label{g8}
8\pi\left(\tilde{\mu}+\frac{s^2}{8\pi
r^4}-\mathcal{T}_{0}^{0(\mathcal{D})}-\zeta
\Upsilon_{0}^{0}\right)&=e^{-\beta}\left(\frac{\beta'}{r}-\frac{1}{r^2}\right)
+\frac{1}{r^2},\\\label{g9} 8\pi\left(\tilde{p}-\frac{s^2}{8\pi
r^4}+\mathcal{T}_{1}^{1(\mathcal{D})}+\zeta\Upsilon_{1}^{1}\right)&=e^{-\beta}\left(\frac{\chi'}{r}+\frac{1}{r^2}\right)
-\frac{1}{r^2},
\\\label{g10}
8\pi\left(\tilde{p}+\frac{s^2}{8\pi
r^4}+\mathcal{T}_{2}^{2(\mathcal{D})}+\zeta\Upsilon_{2}^{2}\right)&=\frac{e^{-\beta}}{4}\left[2\chi''+\chi'^2-\chi'\beta'
+\frac{2\chi'}{r}-\frac{2\beta'}{r}\right],
\end{align}
where $\tilde{\mu}=\frac{1}{(f_{\mathcal{R}}+\mu
f_{\mathcal{Q}})}\mu$,~$\tilde{p}=\frac{1}{(f_{\mathcal{R}}+\mu
f_{\mathcal{Q}})}p$ and $s^2=\frac{1}{(f_{\mathcal{R}}+\mu
f_{\mathcal{Q}})}q^2$. The inclusion of charge as well as modified
corrections
$\mathcal{T}_{0}^{0(\mathcal{D})},~\mathcal{T}_{1}^{1(\mathcal{D})}$
and $\mathcal{T}_{2}^{2(\mathcal{D})}$ produce much complications in
the field equations \eqref{g8}-\eqref{g10}. The values of these
components are presented in Appendix \textbf{A}.

The existence of matter-geometry coupling in this gravitational
theory assures the non-vanishing divergence of stress-energy tensor,
(i.e., $\nabla_\lambda \mathcal{T}^{\lambda\xi}\neq 0$) in contrast
to GR and $f(\mathcal{R})$ theory which results in violation of the
equivalence principle. This violation generates an additional
force in the system due to which the particles moving in the
gravitational field do not obey geodesic path. Therefore we obtain
\begin{align}\nonumber
\nabla^\lambda\big(\mathcal{T}_{\lambda\xi}+\mathcal{E}_{\lambda\xi}+\Upsilon_{\lambda\xi}\big)&=\frac{2}{2f_\mathcal{T}
+\mathcal{R}f_\mathcal{Q}+16\pi}\left[\nabla_\xi(\mathcal{L}_mf_\mathcal{T})
+\nabla_\lambda(f_\mathcal{Q}\mathcal{R}^{\rho\lambda}\mathcal{T}_{\rho\xi})\right.\\\nonumber
&-\frac{1}{2}
(f_\mathcal{T}g_{\rho\eta}+f_\mathcal{Q}\mathcal{R}_{\rho\eta})\nabla_\xi
\mathcal{T}^{\rho\eta}-\mathcal{G}_{\lambda\xi}\nabla^\lambda(f_\mathcal{Q}\mathcal{L}_m)\\\label{g11}
&-\left.\frac{1}{2}\big\{\nabla^{\lambda}(\mathcal{R}f_{\mathcal{Q}})+2\nabla^{\lambda}f_{\mathcal{T}}\big\}\mathcal{T}_{\lambda\xi}\right].
\end{align}
Using the above equation, the condition for hydrostatic equilibrium
becomes
\begin{equation}\label{g12}
\frac{dp}{dr}+\zeta \frac{d\Upsilon_{1}^{1}}{dr}-\frac{ss'}{4\pi
r^4}+\frac{\chi'}{2}\left(\mu+p\right)-\frac{2\zeta}{r}
\left(\Upsilon_{2}^{2}-\Upsilon_{1}^{1} \right)-\frac{\zeta\chi'}{2}
\left(\Upsilon_{0}^{0}-\Upsilon_{1}^{1}\right)=\Omega,
\end{equation}
where $\Omega$ appears due to the condition \eqref{g11}. Its value
is casted in Appendix \textbf{A}. Equation \eqref{g12} can be
pointed out as the generalization of Tolman-Opphenheimer-Volkoff
(TOV) equation. This equation plays a key role in studying the
systematic changes in spherically self-gravitating configurations.

The complex differential equations \eqref{g8}-\eqref{g10} and
\eqref{g12} are found to be highly non-linear involving eight
unknowns
$(\chi,\beta,\mu,p,s,\Upsilon_{0}^{0},\Upsilon_{1}^{1},\Upsilon_{2}^{2})$
which make the system indefinite. Thus we utilize the systematic
strategy \cite{33} to close the system and then calculate the
unknowns. We express the modified physical variables appear in the
field equations \eqref{g8}-\eqref{g10} as
\begin{equation}\label{g13}
\hat{\mu}=\tilde{\mu}-\zeta\Upsilon_{0}^{0},\quad
\hat{p}_{r}=\tilde{p}+\zeta\Upsilon_{1}^{1}, \quad
\hat{p}_{\bot}=\tilde{p}+\zeta\Upsilon_{2}^{2}.
\end{equation}
This indicates that the new source $\Upsilon_{\lambda}^{\xi}$ causes
anisotropy inside a self-gravitating system. We thus define it as
\begin{equation}\label{g14}
\hat{\Pi}=\hat{p}_{\bot}-\hat{p}_{r}=\zeta\left(\Upsilon_{2}^{2}-\Upsilon_{1}^{1}\right),
\end{equation}
which vanishes for $\zeta=0$.

\section{Extended Gravitational Decoupling}

We now work out the system \eqref{g8}-\eqref{g10} to determine
unknown quantities through an innovative algorithm referred to
gravitational decoupling through EGD technique. The effective field
equations are therefore transformed through this method such that
the additional source $\Upsilon_{\lambda}^{\xi}$ may guarantee the
existence of anisotropy in the inner geometry. The key element of
this technique is the ideal fluid solution $(\phi,\psi,\mu,p,s)$, so
let us begin with the metric given as
\begin{equation}\label{g15}
ds^2=-e^{\phi}dt^2+e^{\psi}dr^2+r^2d\theta^2+r^2\sin^2\theta
d\varphi^2,
\end{equation}
where $\phi=\phi(r)$ and
$\psi=\psi(r)=1-\frac{2m(r)}{r}+\frac{s^2}{r^2}$. Here, $m(r)$
indicates the Misner-Sharp mass of the spherical distribution
\eqref{g6}. Further, we reconstruct the metric potentials by taking
two geometrical transformations on them and evaluate the influence
of a source $\Upsilon_{\lambda}^{\xi}$ on isotropic solution in the
presence of charge. Thus the transformations are
\begin{equation}\label{g16}
\phi\rightarrow\chi=\phi+\zeta \mathfrak{l}, \quad
e^{-\psi}\rightarrow e^{-\beta}=e^{-\psi}+\zeta \mathfrak{n},
\end{equation}
where $\mathfrak{l}=\mathfrak{l}(r)$ and
$\mathfrak{n}=\mathfrak{n}(r)$ confirm their correspondence with
temporal and radial metric functions, respectively. Therefore, EGD
technique ensures that both components are translated.

We require the solution of complex field equations, thus for our
ease, we divide them into two different sets by imposing the
overhead transformations on system \eqref{g8}-\eqref{g10}. For
$\zeta=0$, the first set takes the form as
\begin{align}\label{g18}
&8\pi\left(\tilde{\mu}+\frac{s^2}{8\pi
r^4}-\mathcal{T}_{0}^{0(\mathcal{D})}\right)=\frac{1}{r^2}+e^{-\psi}\left(\frac{\psi'}{r}-\frac{1}{r^2}\right),\\\label{g19}
&8\pi\left(\tilde{p}-\frac{s^2}{8\pi
r^4}+\mathcal{T}_{1}^{1(\mathcal{D})}\right)=-\frac{1}{r^2}+e^{-\psi}\left(\frac{\phi'}{r}+\frac{1}{r^2}\right),\\\label{g20}
&8\pi\left(\tilde{p}+\frac{s^2}{8\pi
r^4}+\mathcal{T}_{2}^{2(\mathcal{D})}\right)=e^{-\psi}\left(\frac{\phi''}{2}+\frac{\phi'^2}{4}-\frac{\phi'\psi'}{4}
+\frac{\phi'}{2r}-\frac{\psi'}{2r}\right),
\end{align}
and the second set which engages the source
$\Upsilon^{\xi}_{\lambda}$ as well as transformation functions
becomes
\begin{align}\label{g21}
8\pi\Upsilon_{0}^{0}&=\frac{\mathfrak{n}'}{r}+\frac{\mathfrak{n}}{r^2},\\\label{g22}
8\pi\Upsilon_{1}^{1}&=\mathfrak{n}\left(\frac{\chi'}{r}+\frac{1}{r^2}\right)+\frac{e^{-\psi}\mathfrak{l}'}{r},\\\nonumber
8\pi\Upsilon_{2}^{2}&=\frac{\mathfrak{n}}{4}\left(2\chi''+\chi'^2+\frac{2\chi'}{r}\right)+\frac{e^{-\psi}}{4}\left(2\mathfrak{l}''
+\zeta\mathfrak{l}'^2+\frac{2\mathfrak{l}'}{r}+2\phi'\mathfrak{l}'-\psi'\mathfrak{l}'\right)\\\label{g23}
&+\frac{\mathfrak{n}'}{4}\left(\chi' +\frac{2}{r}\right).
\end{align}
The field equations for ideal fluid configuration vary from
Eqs.\eqref{g21}-\eqref{g23} only by few terms. These equations can
therefore be marked as the typical anisotropic field equations
corresponding to spherical spacetime stated as
\begin{equation}\label{g24}
\Upsilon^{*\xi}_{\lambda}=\Upsilon^{\xi}_{\lambda}-\frac{1}{r^2}\delta_{\lambda}^{0}\delta_{0}^{\xi}
-\left(\mathcal{A}_{1}+\frac{1}{r^2}\right)\delta_{\lambda}^{1}\delta_{1}^{\xi}-\mathcal{A}_{2}\delta_{\lambda}^{2}\delta_{2}^{\xi},
\end{equation}
with precise notations
\begin{align}\label{g24a}
\Upsilon^{*0}_{0}&=\Upsilon^{0}_{0}-\frac{1}{r^2},\\\label{g24b}
\Upsilon^{*1}_{1}&=\Upsilon^{1}_{1}-\left(\mathcal{A}_{1}+\frac{1}{r^2}\right),\\\label{g24c}
\Upsilon^{*2}_{2}&=\Upsilon^{2}_{2}-\mathcal{A}_{2},
\end{align}
where
\begin{equation}\nonumber
\mathcal{A}_{1}=\frac{e^{-\psi}\mathfrak{l}'}{r}, \quad
\mathcal{A}_{2}=\frac{e^{-\psi}}{4}\left(2\mathfrak{l}''
+\zeta\mathfrak{l}'^2+\frac{2\mathfrak{l}'}{r}+2\phi'\mathfrak{l}'-\psi'\mathfrak{l}'\right).
\end{equation}
As a consequence, an indefinite system \eqref{g8}-\eqref{g10} has
been divided into two sectors in which the first set
\eqref{g18}-\eqref{g20} exhibits the equations of motion for
isotropic fluid $(\tilde{\mu},\tilde{p},\chi,\beta$). It is observed
that the second sector \eqref{g21}-\eqref{g23} obeys the anisotropic
system \eqref{g24} involving five unknowns
($\mathfrak{l},\mathfrak{n},\Upsilon_{0}^{0},\Upsilon_{1}^{1},\Upsilon_{2}^{2}$).
Eventually, we have decoupled the system \eqref{g8}-\eqref{g10}
successfully.

Several constraints on the boundary surface $(\Sigma)$ play a vital
role to explore basic characteristics of massive structures. These
constraints are termed as junction conditions which help us to match
the inner and outer regions of the compact object at the boundary.
Thus the interior geometry is taken as
\begin{equation}\label{g25}
ds^2=-e^{\chi}dt^2+\frac{1}{\left(1-\frac{2m}{r}+\frac{s^2}{r^2}+\zeta\mathfrak{n}\right)}dr^2+
r^2d\theta^2+r^2\sin^2\theta d\varphi^2.
\end{equation}
We take exterior spacetime corresponding to the geometry \eqref{g6}
so that we can match it smoothly with the interior geometry
\eqref{g25}. The first fundamental form of the junction conditions
ensures the equivalence of both inner and outer geometries at the
hypersurface, i.e., $([ds^2]_{\Sigma}=0)$ yields
\begin{equation}\label{g27}
\phi+\zeta \mathfrak{l}(H)=\chi_{-}(H)=\chi_{+}(H), \quad
1-e^{-\beta_{+}(H)}=\frac{2\mathcal{M}}{H}-\frac{\mathcal{S}^2}{H^2}-\zeta\mathfrak{n}(H),
\end{equation}
where the symbols $-$ and $+$ indicate that the metric components
correspond to the inner and outer spacetimes, respectively.
Moreover,
$\mathcal{M}=m(H),~\mathcal{S}^2=s^2(H)$,~$\mathfrak{l}(H)$ and
$\mathfrak{n}(H)$ define the total mass, charge and deformation
functions of spherical body at the boundary. Likewise, the second
form
\big($[\mathcal{T}^{(tot)}_{\lambda\xi}\mathcal{W}^{\xi}]_{\Sigma}=0$,
where $\mathcal{W}^\xi=(0,e^{\frac{-\beta}{2}},0,0)$ is the
four-vector\big) delivers
\begin{equation}\label{g28}
\tilde{p}(H)-\frac{\mathcal{S}^{2}}{8\pi H^4}+\zeta
\left(\Upsilon^{1}_{1}(H)\right)_{-}+
\left(\mathcal{T}^{1(\mathcal{D})}_{1}(H)\right)_{-}=\zeta
\left(\Upsilon^{1}_{1}(H)\right)_{+}+\left(\mathcal{T}^{1(\mathcal{D})}_{1}(H)\right)_{+}.
\end{equation}
The above equation takes the form after using Eq.\eqref{g27} as
\begin{equation}\label{g29}
\tilde{p}(H)-\frac{\mathcal{S}^{2}}{8\pi H^4}+\zeta
\left(\Upsilon^{1}_{1}(H)\right)_{-}=\zeta
\left(\Upsilon^{1}_{1}(H)\right)_{+},
\end{equation}
which can further be expressed as
\begin{eqnarray}\nonumber
&&\tilde{p}(H)-\frac{\mathcal{S}^{2}}{8\pi
H^4}+\frac{\zeta}{8\pi}\left[\mathfrak{n}(H)\left(\frac{\chi'(H)}{H}+
\frac{1}{H^2}\right)+e^{-\psi_{H}}\frac{\mathfrak{l}'_{H}}{H}\right]=\frac{\zeta}{8\pi}\bigg[\mathfrak{n}^*(H)\\\label{g30}
&&\times\left\{\frac{1}{H^2}
+\frac{1}{H^2}\left(\frac{2\bar{M}H-2\bar{S}^2}{H^2-2\bar{M}H+\bar{S}^2}\right)\right\}+e^{-\psi_{H}}\frac{\mathfrak{l^*}'_{H}}{H}\bigg].
\end{eqnarray}
The term $\bar{M}$ presents the mass,~$\bar{S}$ is the charge and
$\mathfrak{l}^*$ as well as $\mathfrak{n}^*$ are the temporal and
radial deformations of the outer Reissner-Nordstr\"{o}m geometry
which is affected by $\Upsilon_{\lambda\xi}$ (source). Hence the
metric is described as
\begin{eqnarray}\nonumber
ds^2&=&-\left(1-\frac{2\bar{M}}{r}+\frac{\bar{S}^2}{r^2}+\zeta\mathfrak{l}^*(r)\right)dt^2+\frac{1}{\left(1-\frac{2\bar{M}}{r}+\frac{\bar{S}^2}{r^2}
+\zeta\mathfrak{n}^*\right)}dr^2\\\label{g31}
&+&r^2d\theta^2+r^2\sin^2\theta d\varphi^2.
\end{eqnarray}
Equations \eqref{g27} and \eqref{g30} provide certain suitable
conditions which interlink the EGD inner spherical structure with
outer Reissner-Nordstr\"{o}m geometry, both of which are filled with
source $(\Upsilon_{\lambda\xi})$.

\section{Anisotropic Solutions}

Our goal is to construct two anisotropic solutions with the help of
EGD approach and utilize different constraints to close the system.
To make this happen, we require isotropic solution of the field
equations \eqref{g18}-\eqref{g20} in
$f(\mathcal{R},\mathcal{T},\mathcal{Q})$ scenario. Thus we continue
our study by considering the non-singular Krori-Barua solution for
isotropic configuration \cite{42} which takes the form in this
gravity as
\begin{eqnarray}\label{g33}
e^{\chi}&=&e^{Ar^2+C}, \\\label{g34} e^{\beta}&=&e^{\psi}=e^{Br^2},
\\\nonumber \tilde{\mu}&=&\frac{1}{16\pi}\left[e^{-Br^2}
\left\{5B-Ar^2(A-B)-\frac{1}{r^2}\right\}+\frac{1}{r^2}\right]
+\mathcal{T}^{0(\mathcal{D})}_{0}-\frac{1}{2}\\\label{g35}
&\times&\left(\mathcal{T}^{1(\mathcal{D})}_{1}-\mathcal{T}^{2(\mathcal{D})}_{2}\right),\\\nonumber
\tilde{p}&=&\frac{1}{16\pi}\left[e^{-Br^2}\left\{4A-B+Ar^2(A-B)
+\frac{1}{r^2}\right\}-\frac{1}{r^2}\right]-\frac{1}{2}\\\label{g36}
&\times&\left(\mathcal{T}^{1(\mathcal{D})}_{1}+\mathcal{T}^{2(\mathcal{D})}_{2}\right),\\\nonumber
s^2&=&-\frac{r^2}{2}\left[e^{-Br^2}\left\{1+Br^2+Ar^4(B-A)
\right\}-1\right]+4\pi r^4\\\label{g36a}
&\times&\left(\mathcal{T}^{1(\mathcal{D})}_{1}-\mathcal{T}^{2(\mathcal{D})}_{2}\right).
\end{eqnarray}
The above set of equations exhibits three constant $A,~B$ and $C$ as
unknowns whose values are calculated at the boundary $r=H$ by
employing continuity of metric functions $(g_{tt},~g_{rr}$ and
$g_{tt,r})$ as
\begin{eqnarray}\label{g37}
A&=&\frac{1}{H^2}\left(\frac{\mathcal{M}}{H}-\frac{\mathcal{S}^{2}}{H^2}\right)
\left(1-\frac{2\mathcal{M}}{H}+\frac{\mathcal{S}^{2}}{H^2}\right)^{-1},\\\label{g38}
B&=&-\frac{1}{H^2}
\ln\left(1-\frac{2\mathcal{M}}{H}+\frac{\mathcal{S}^{2}}{H^2}\right),\\\label{g38a}
C&=&\ln\left(1-\frac{2\mathcal{M}}{H}+\frac{\mathcal{S}^{2}}{H^2}\right)
-\frac{\mathcal{M}H-\mathcal{S}^{2}}{H^2-2\mathcal{M}H+\mathcal{S}^{2}},
\end{eqnarray}
where compactness factor is defined as
$\frac{2\mathcal{M}}{H}<\frac{8}{9}$. The compatibility of interior
isotropic solution \eqref{g33}-\eqref{g36a} with the exterior
Reissner-Nordstr\"{o}m geometry is pledged by
Eqs.\eqref{g37}-\eqref{g38a} at the boundary $(r=H)$, that may be
modified in the interior due to the incorporation of additional
source $\Upsilon_{\lambda\xi}$. The anisotropic solutions for inner
spacetime can be developed by utilizing the temporal and radial
metric components in terms of Krori-Barua ansatz \eqref{g33} and
\eqref{g34}. Equations \eqref{g21}-\eqref{g23} connect the source
$\Upsilon_{\lambda\xi}$ with geometric deformations $\mathfrak{l}$
and $\mathfrak{n}$ in an interesting way and we determine their
solution by making use of certain conditions.

In the following, some constraints are considered to find two
anisotropic charged solutions and we check their feasibility as well
through graphical behavior.

\subsection{Solution I}

Here, we employ a linear equation of state to calculate first
anisotropic solution as
\begin{equation}\label{g38b}
\Upsilon^{0}_{0}=\tau\Upsilon^{1}_{1}+\upsilon\Upsilon^{2}_{2}.
\end{equation}
We consider another constraint on $\Upsilon_{1}^{1}$ to calculate
$\mathfrak{l},~\mathfrak{n}$ and $\Upsilon_{\lambda}^{\xi}$. We set
$\tau=1$ and $\upsilon=0$ for our convenience, thus the relation
\eqref{g38b} takes the form $\Upsilon^{0}_{0}=\Upsilon^{1}_{1}$. We
take the constraint $\tilde{p}(H)-\frac{\mathcal{S}^2}{8\pi
H^4}+\mathcal{T}_{1}^{1(\mathcal{D})}(H)\sim -\zeta
\left(\Upsilon_{1}^{1}(H)\right)_{-}$ which assures the
compatibility between interior isotropic composition and exterior
Reissner-Nordstr\"{o}m. The forthright choice which meets this
requirement is \cite{33}
\begin{equation}\label{g39}
-\tilde{p}+\frac{s^2}{8\pi
r^4}-\mathcal{T}_{1}^{1(\mathcal{D})}=\Upsilon_{1}^{1}.
\end{equation}
After using the field equations \eqref{g19},~\eqref{g21} and
\eqref{g22} in constraints \eqref{g38b} and \eqref{g39} we have
deformation functions as
\begin{eqnarray}\label{g39a}
\mathfrak{l}(r)&=&\int
\frac{1-\left(\mathfrak{n}(r)+e^{\beta}\right)\left(r\chi'
e^{\chi}+1\right)}{r\left(\zeta\mathfrak{n}(r)+e^{\beta}\right)}dr,\\\label{g39b}
\mathfrak{n}(r)&=&1-\frac{1}{r}\int e^{\beta}\left(r\chi'
e^{\chi}+1\right)dr,
\end{eqnarray}
which take the form in terms of Krori-Barua ansatz \eqref{g33} and
\eqref{g34} as
\begin{eqnarray}\nonumber
\mathfrak{l}&=&\int \frac{1}{r\varpi}\left[\sqrt{\pi}e^{B
r^2}(A+B)\left(2Ar^2+1\right)\text{erf}\left(\sqrt{B}r\right)-2
\sqrt{B}r\left\{A\left(2Ar^2\right.\right.\right.\\\label{g39c}
&+&\left.\left.\left.1\right)+B\left(2Ar^2\left(e^{Br^2}+1\right)+1\right)\right\}\right]dr,\\\label{g39d}
\mathfrak{n}&=&1-\frac{\sqrt{\pi}\left(A+B\right)
\text{erf}\left(\sqrt{B}r\right)}{2B^{3/2}r}+\frac{Ae^{-Br^2}}{B},
\end{eqnarray}
where
\begin{equation}\nonumber
\varpi=2\sqrt{B}r\left(B\zeta e^{Br^2}+B+A\zeta\right)-\sqrt{\pi
}\zeta\left(A+B\right)e^{Br^2}\text{erf}\left(\sqrt{B}r\right).
\end{equation}
The temporal and radial components thus become
\begin{eqnarray}\nonumber
\chi&=&Ar^2+C+\zeta \int\frac{1}{r\varpi}\left[\sqrt{\pi}e^{B
r^2}(A+B)\left(2Ar^2+1\right)\text{erf}\left(\sqrt{B}r\right)\right.\\\label{g40}
&-&\left.2 \sqrt{B}r\left\{A\left(2Ar^2+1\right)+B\left(2A
r^2\left(e^{Br^2}+1\right)+1\right)\right\}\right]dr,\\\label{g40a}
e^{-\beta}&=&e^{-Br^2}+\zeta\left(1-\frac{\sqrt{\pi}\left(A+B\right)
\text{erf}\left(\sqrt{B}r\right)}{2B^{3/2}r}+\frac{Ae^{-Br^2}}{B}\right).
\end{eqnarray}
The above expressions will reduce to the standard Krori-Barua
solution corresponding to the ideal spherical fluid $(\zeta=0)$.

We utilize the matching criteria at hypersurface $\Sigma$ to
investigate the impact of pressure anisotropy on triplet $(A,B,C)$.
We also obtain the following relations from first fundamental form
of junction conditions as
\begin{eqnarray}\nonumber
\ln\left(1-\frac{2\mathcal{M}}{H}+\frac{\mathcal{S}^2}{H^2}\right)&=&AH^2+C+\zeta
\left[\int\frac{1}{r\varpi}\left[\sqrt{\pi}e^{B
r^2}(A+B)\left(2Ar^2+1\right)\right.\right.\\\nonumber
&\times&\text{erf}\left(\sqrt{B}r\right)-2\sqrt{B}r\left\{B\left(2
Ar^2\left(e^{Br^2}+1\right)+1\right)\right.\\\label{g41}
&+&\left.\left.\left.A\left(2Ar^2+1\right)\right\}\right]dr\right]_{r=H},
\end{eqnarray}
and
\begin{eqnarray}\label{g42}
1-\frac{2\mathcal{M}}{H}+\frac{\mathcal{S}^2}{H^2}=e^{-BH^2}+\zeta\left(1-\frac{\sqrt{\pi}\left(A+B\right)
\text{erf}\left(\sqrt{B}H\right)}{2B^{3/2}H}+\frac{Ae^{-BH^2}}{B}\right).
\end{eqnarray}
On the other hand, the second fundamental form
$(\tilde{p}(H)-\frac{S^2}{8\pi
H^4}+\mathcal{T}_{1}^{1(\mathcal{D})}(H)+\zeta
\left(\Upsilon_{1}^{1}(H)\right)_{-}=0)$ yields
\begin{equation}\label{g43}
\tilde{p}(H)-\frac{s^2}{8\pi
r^4}+\mathcal{T}_{1}^{1(\mathcal{D})}(H)=0 \quad\Rightarrow\quad
B=\frac{\ln\left(1+2AH^2\right)}{H^2}.
\end{equation}
Equation \eqref{g43} shows the relation between two constants $A$
and $B$. Equations \eqref{g41}-\eqref{g43} supply certain suitable
conditions through which we can smoothly match both regions of
spherical geometry at the boundary. Finally, the constraints
\eqref{g38b} and \eqref{g39} together with the field equations as
well as Eq.\eqref{g13} produce the following anisotropic solution in
the presence of charge as
\begin{align}\nonumber
\hat{\mu}&=\frac{1}{16\pi}\left[e^{-Br^2}\left\{B\left(5+Ar^2\right)+A\left(4\zeta-Ar^2\right)-\frac{1}{r^2}\left(1-2\zeta\right)\right\}
+\frac{1}{r^2}\right.\\\label{g46}
&\times\left.(1-2\zeta)+16\pi\mathcal{T}^{0(\mathcal{D})}_{0}
-8\pi\left(\mathcal{T}^{1(\mathcal{D})}_{1}-\mathcal{T}^{2(\mathcal{D})}_{2}\right)\right],\\\nonumber
\hat{p}_{r}&=\frac{1}{16\pi}\left[e^{-Br^2}\left\{A\left(4+Ar^2-4\zeta\right)-B\left(1+Ar^2\right)+\frac{1}{r^2}\left(1-2\zeta\right)\right\}
-\frac{1}{r^2}\right.\\\label{g47}
&\times\left.(1-2\zeta)-8\pi\left(\mathcal{T}^{1(\mathcal{D})}_{1}+\mathcal{T}^{2(\mathcal{D})}_{2}\right)\right],\\\nonumber
\hat{p}_{\bot}&=\frac{1}{8\pi}\left[\frac{e^{-Br^2}}{2}\left(Ar^2(A-B)-B+4A+\frac{1}{r^2}\right)-\frac{1}{2r^2}+\frac{\zeta}{4B^{3/2}
r^3}e^{-Br^2} \left(Ar^2\right.\right.\\\nonumber
&+\left.1\right)\left(\sqrt{\pi } (A+B) e^{Br^2}
\text{erf}\left(\sqrt{B}r\right)-2 \sqrt{B}r\left(2 A B
r^2+A+B\right)\right)+\zeta A\\\nonumber &\times\left(Ar^2+2\right)
\left(1-\frac{\sqrt{\pi } (A+B) \text{erf}\left(\sqrt{B} r\right)}{2
B^{3/2} r}+\frac{Ae^{-Br^2}}{B}\right)+\frac{\zeta}{4\varpi}\left\{2
Be^{-Br^2}\right.\\\nonumber &\times\left(\sqrt{\pi} (A+B) e^{Br^2}
\left(2Ar^2+1\right) \text{erf}\left(\sqrt{B}r\right)-2\sqrt{B}r
\left(B\left(2Ar^2
\left(e^{Br^2}+1\right)\right.\right.\right.\\\nonumber
&+\left.\left.\left.\left.1\right)+A\left(2Ar^2+1\right)\right)\right)\right\}+\frac{e^{-Br^2}}{4r^2\varpi^2}\left\{4
Br^2 \left(4B^3r^2e^{Br^2} \left(\zeta +2(\zeta
-1)Ar^2\right)\right.\right.\\\nonumber &-B^2 \left(4A^2r^4
\left(e^{Br^2}+1\right) \left(\zeta
\left(e^{Br^2}-1\right)+2\right)+4Ar^2 \left(e^{Br^2}(\zeta
+2)+2\right.\right.\\\nonumber &+\left.\left.2 \zeta
e^{2Br^2}\right)-2 \zeta e^{Br^2}+\zeta -2\right)-2AB\left(4A^2r^4
\left(\zeta  e^{Br^2}+1\right)+4Ar^2\right.\\\nonumber
&\times\left.\left. \left(2 \zeta  e^{Br^2}+\zeta +1\right)-\zeta
e^{Br^2}-1\right)+\zeta A^2 \left(-4A^2r^4-8Ar^2+1\right)\right)-4
\sqrt{\pi}\\\nonumber &\times\sqrt{B}r(A+B)e^{Br^2}
\text{erf}\left(\sqrt{B} r\right)\left(2B^2 (\zeta -1) r^2
\left(2Ar^2+1\right)-B\left(4A^2 r^4\right.\right.\\\nonumber
&\times\left.\left(\zeta e^{Br^2}+1\right)+4Ar^2 \left(2\zeta
e^{Br^2}+\zeta+1\right)-\zeta e^{Br^2}-1\right)+\zeta A\left(-4A^2
r^4\right.\\\nonumber &-\left.\left.\left.8Ar^2+1\right)\right)-\pi
\zeta (A+B)^2 e^{2Br^2} \left(4A^2r^4+8Ar^2-1\right)
\text{erf}\left(\sqrt{B}r\right)^2\right\}\\\label{g48}
&-\left.4\pi\left(\mathcal{T}^{1(\mathcal{D})}_{1}+\mathcal{T}^{2(\mathcal{D})}_{2}\right)\right],\\\nonumber
s^2&=\frac{r^2}{2}-\frac{1}{2} e^{-Br^2} \left(ABr^6+Br^4-A^2
r^6+r^2\right)-2\pi r^2(1-2\zeta)+2\pi r^4e^{-Br^2}\\\nonumber
&\times\left(-B\left(Ar^2+1\right)+A\left(-4 \zeta
+Ar^2+4\right)+\frac{1-2 \zeta}{r^2}\right)-2\pi r^2+2\pi
r^4\\\nonumber &\times e^{-Br^2} \left(Ar^2(A-B)-B+4
A+\frac{1}{r^2}\right)+\frac{\zeta\pi r^4}{B^{3/2}r^3}e^{-Br^2}
\left(Ar^2+1\right)\\\nonumber &\times\left(\sqrt{\pi} (A+B)
e^{Br^2}\text{erf}\left(\sqrt{B} r\right)-2 \sqrt{B} r \left(2AB
r^2+A+B\right)\right)+4\pi Ar^4\\\nonumber &\times\left(A
r^2+2\right)\left(1-\frac{\sqrt{\pi}(A+B)\text{erf}\left(\sqrt{B}
r\right)}{2B^{3/2} r}+\frac{Ae^{-Br^2}}{B}\right)+\frac{2\pi B\zeta
r^4e^{-Br^2}}{\varpi}\\\nonumber
&\times\left\{\sqrt{\pi}(A+B)e^{Br^2} \left(2Ar^2+1\right)
\text{erf}\left(\sqrt{B}r\right)-2 \sqrt{B}r\left(B\left(2Ar^2
\left(e^{Br^2}+1\right)\right.\right.\right.\\\nonumber
&+\left.\left.\left.1\right)+A\left(2Ar^2+1\right)\right)\right\}+\frac{e^{-Br^2}}{r^2
\varpi^2}\left[4Br^2 \left(4B^3 r^2e^{Br^2}\left(\zeta +2(\zeta -1)
Ar^2\right)\right.\right.\\\nonumber &-B^2
\left(4A^2r^4\left(e^{Br^2}+1\right)\left(\zeta\left(e^{Br^2}-1\right)+2\right)+4
Ar^2 \left(2\zeta e^{2Br^2}+(\zeta +2)\right.\right.\\\nonumber
&\times\left.\left. e^{Br^2}+2\right)-2\zeta e^{Br^2}+\zeta
-2\right)-2AB\left(4A^2r^4\left(\zeta e^{Br^2}+1\right)-\zeta
e^{Br^2}-1\right.\\\nonumber &+\left.4Ar^2 \left(2\zeta
e^{Br^2}+\zeta +1\right)\right)+\zeta A^2 \left(-4A^2
r^4-8Ar^2+1\right)-4B^3r^2 \left(4A^2r^4\right.\\\nonumber
&\times\left(e^{Br^2}+1\right)\left(\zeta
\left(e^{Br^2}-1\right)+2\right)+4Ar^2 \left(2\zeta e^{2Br^2}+(\zeta
+2)e^{Br^2}+2\right)\\\nonumber &-\left.2 \zeta e^{Br^2}+\zeta
-2\right)-4\sqrt{\pi} \sqrt{B}r(A+B) e^{Br^2}
\text{erf}\left(\sqrt{B} r\right) \left(2B^2 r^2
\left(2Ar^2+1\right)\right.\\\nonumber &\times (\zeta-1)-B\left(4A^2
r^4 \left(\zeta e^{Br^2}+1\right)+4Ar^2 \left(2\zeta e^{Br^2}+\zeta
+1\right)-\zeta e^{Br^2}-1\right)\\\nonumber &+\left.\zeta
A\left(-4A^2r^4-8Ar^2+1\right)\right)-\pi \zeta (A+B)^2 e^{2Br^2}
\left(4A^2r^4+8A r^2-1\right)\\\label{48a}
&\times\left.\left.\text{erf}\left(\sqrt{B}
r\right)^2\right)\right]+4\pi
r^4\left(\mathcal{T}^{1(\mathcal{D})}_{1}
-\mathcal{T}^{2(\mathcal{D})}_{2}\right),
\end{align}
and the pressure anisotropy is
\begin{align}\nonumber
\hat{\Pi}&=\frac{1}{8\pi}\left[\frac{e^{-Br^2}}{2}\left(Ar^2(A-B)-B+4A+\frac{1}{r^2}\right)-\frac{1}{2r^2}+\frac{\zeta}{4B^{3/2}
r^3}e^{-Br^2} \left(Ar^2\right.\right.\\\nonumber
&+\left.1\right)\left(\sqrt{\pi} (A+B) e^{Br^2}
\text{erf}\left(\sqrt{B} r\right)-2 \sqrt{B} r
\left(2ABr^2+A+B\right)\right)+\zeta A\\\nonumber
&\times\left(Ar^2+2\right) \left(1-\frac{\sqrt{\pi } (A+B)
\text{erf}\left(\sqrt{B}
r\right)}{2B^{3/2}r}+\frac{Ae^{-Br^2}}{B}\right)+\frac{\zeta}{4\varpi}\left\{2Be^{-A
r^2}\right.\\\nonumber &\times\left(\sqrt{\pi} (A+B) e^{Br^2}
\left(2Ar^2+1\right) \text{erf}\left(\sqrt{B} r\right)-2 \sqrt{B} r
\left(B\left(2Ar^2
\left(e^{Br^2}+1\right)\right.\right.\right.\\\nonumber
&+\left.\left.\left.\left.1\right)+A\left(2Ar^2+1\right)\right)\right)\right\}+\frac{e^{-Br^2}}{4r^2\varpi^2}\left\{4Br^2
\left(4B^3 r^2 e^{Br^2} \left(\zeta +2(\zeta
-1)Ar^2\right)\right.\right.\\\nonumber &-B^2 \left(4A^2r^4
\left(e^{Br^2}+1\right)
\left(\zeta\left(e^{Br^2}-1\right)+2\right)+4A r^2
\left(e^{Br^2}(\zeta +2)+2\right.\right.\\\nonumber &+\left.\left.2
\zeta e^{2Br^2}\right)-2\zeta e^{Br^2}+\zeta-2\right)-2AB\left(4A^2
r^4 \left(\zeta e^{Br^2}+1\right)+4Ar^2\right.\\\nonumber
&\times\left.\left. \left(2\zeta e^{Br^2}+\zeta +1\right)-\zeta
e^{Br^2}-1\right)+\zeta A^2 \left(-4A^2r^4-8Ar^2+1\right)\right)-4
\sqrt{\pi}\\\nonumber &\times\sqrt{B}r(A+B) e^{Br^2}
\text{erf}\left(\sqrt{B} r\right)\left(2B^2(\zeta -1)r^2
\left(2Ar^2+1\right)-B\left(4A^2 r^4\right.\right.\\\nonumber
&\times\left.\left(\zeta e^{Br^2}+1\right)+4Ar^2 \left(2\zeta
e^{Br^2}+\zeta +1\right)-\zeta e^{Br^2}-1\right)+\zeta A\left(-4A^2
r^4\right.\\\nonumber &-\left.\left.\left.8Ar^2+1\right)\right)-\pi
\zeta (A+B)^2 e^{2Br^2} \left(4A^2 r^4+8Ar^2-1\right)
\text{erf}\left(\sqrt{B} r\right)^2\right\}\\\nonumber
&-\frac{e^{-Br^2}}{2}\left\{A\left(4+Ar^2-4\zeta\right)-B\left(1+Ar^2\right)+\frac{1}{r^2}\left(1-2\zeta\right)\right\}
+\frac{1}{2r^2}\\\label{g49}
&\times\left.\left(1-2\zeta\right)\right].
\end{align}

\subsection{Solution II}

To determine another solution for the modified field equations
involving anisotropic source, we take density-like restraint as
\begin{equation}\label{g51}
\tilde{\mu}+\frac{s^2}{8\pi
r^4}-\mathcal{T}_{0}^{0(\mathcal{D})}=\Upsilon_{0}^{0}.
\end{equation}
Combining the field equations \eqref{g18},~\eqref{g21} and
\eqref{g22} with Eqs.\eqref{g38b} and \eqref{g51}, we get
\begin{eqnarray}\label{g51a}
\mathfrak{l}&=&-\int \frac{\beta
e^{\beta}+\left(1-e^{\beta}\right)\chi'e^{\chi}}{\zeta\left(1-e^{\beta}\right)+e^{\beta}}dr,\\\label{g51b}
\mathfrak{n}&=&1-e^{\beta},
\end{eqnarray}
which can also be written together with Eqs.\eqref{g33} and
\eqref{g34} as
\begin{eqnarray}\label{g52}
\mathfrak{l}&=&\int\frac{2r\left(-Ae^{Br^2}+A+B\right)}{\zeta\left(e^{Br^2}-1\right)+1}dr,\\\label{g53}
\mathfrak{n}&=&1-e^{Br^2}.
\end{eqnarray}
Moreover, the matching conditions become for this solution as
\begin{align}\label{g55}
\ln\left(1-\frac{2\mathcal{M}}{H}+\frac{\mathcal{S}^2}{H^2}\right)&=AH^2+C+\zeta\left[\int\frac{2r\left(-Ae^{Br^2}
+A+B\right)}{\zeta\left(e^{Br^2}-1\right)+1}dr\right]_{r=H},\\\label{g56}
B&=-\frac{1}{H^2}\ln\left[1-\frac{1}{1-\zeta}\left(\frac{2\mathcal{M}}{H}-\frac{\mathcal{S}^2}{H^2}\right)\right].
\end{align}
Finally, we formulate the charged anisotropic solution (such as
matter variables and anisotropic factor) for constraints
\eqref{g38b} and \eqref{g51} as
\begin{align}\nonumber
\hat{\mu}&=\frac{1}{16\pi}\bigg[e^{-Br^2}\left\{B\left(5-4\zeta\right)-Ar^2\left(A-B\right)-\frac{1}{r^2}\left(1-2\zeta\right)\right\}
+\frac{1}{r^2}(1-2\zeta)\\\label{g57}
&+16\pi\mathcal{T}^{0(\mathcal{D})}_{0}
-8\pi\left(\mathcal{T}^{1(\mathcal{D})}_{1}-\mathcal{T}^{2(\mathcal{D})}_{2}\right)\bigg],\\\nonumber
\hat{p}_{r}&=\frac{1}{16\pi}\bigg[e^{-Br^2}\left\{A\left(4+Ar^2\right)+B\left(4\zeta-1-Ar^2\right)+\frac{1}{r^2}\left(1-2\zeta\right)\right\}
-\frac{1}{r^2}\\\label{g58} &\times(1-2\zeta)-
8\pi\left(\mathcal{T}^{1(\mathcal{D})}_{1}+\mathcal{T}^{2(\mathcal{D})}_{2}\right)\bigg],\\\nonumber
\hat{p}_{\bot}&=\frac{e^{-Br^2}}{16\pi r^2 \left(\zeta
\left(e^{Br^2}-1\right)+1\right)^2}\bigg[-e^{Br^2}\left(\zeta^2
\left(2B^2r^4-10B\left(Ar^4+r^2\right)+6A^2
r^4\right.\right.\\\nonumber &+\left.16Ar^2+3\right)+2 \zeta ^3 r^2
\left(2B\left(Ar^2+1\right)-3A\left(Ar^2+2\right)\right)+2
\zeta\left(\left(4Ar^4+r^2\right)\right.\\\nonumber
&\times\left.\left.B-4Ar^2-2\right)+1\right)+2B^2\zeta r^4+\zeta ^2
e^{3Br^2}\left(2\zeta A^2r^4+4 \zeta Ar^2-1\right)+\zeta\\\nonumber
&\times e^{2Br^2}\left(\zeta \left(-B \left(3Ar^4+r^2\right)+3A^2
r^4+8Ar^2+3\right)+2\zeta^2r^2
\left(ABr^2+B\right.\right.\\\nonumber
&-\left.\left.3A\left(Ar^2+2\right)\right)-2\right)+Br^2 \left(2
\zeta^3-9 \zeta^2+8 \zeta+\left(2\zeta^3-7\zeta^2+10\zeta
-1\right)\right.\\\nonumber &\times\left.Ar^2-1\right)+\zeta^2-2
\zeta -2 \zeta ^3A^2 r^4+3 \zeta ^2A^2 r^4+A^2r^4-4\zeta^3Ar^2+8
\zeta^2Ar^2\\\label{g59} &-8 \zeta
Ar^2+4Ar^2+1\bigg]-\frac{1}{2}\left(\mathcal{T}^{1(\mathcal{D})}_{1}+\mathcal{T}^{2(\mathcal{D})}_{2}\right),\\\nonumber
s^2&=\frac{r^2}{2}\bigg[4\pi e^{-Br^2}\left(1-Br^2
\left(-4\zeta+Ar^2+1\right)+(2\zeta-1) e^{Br^2}-2\zeta+A^2
r^4\right.\\\nonumber &+\left.4Ar^2\right)-e^{-Br^2}
\left(Ar^2+1\right) \left(Br^2-B r^2+1\right)+1\bigg]-\frac{2\pi
r^2e^{-Br^2}}{\left(\zeta\left(e^{Br^2}-1\right)+1\right)^2}\\\nonumber
&\times\bigg[-e^{Br^2} \left(\zeta^2
\left(2B^2r^4-10B\left(Ar^4+r^2\right)+6A^2r^4+16Ar^2+3\right)+2
\zeta^3 r^2\right.\\\nonumber
&\left.\times\left(2B\left(Ar^2+1\right)-3A\left(Ar^2+2\right)\right)+2
\zeta \left(B\left(4Ar^4+r^2\right)-4A
r^2-2\right)+1\right)\\\nonumber &+2B^2 \zeta r^4+\zeta^2 e^{3Br^2}
\left(2\zeta A^2r^4+4 \zeta Ar^2-1\right)+\zeta e^{2Br^2}
\left(\zeta \left(-B\left(3Ar^4+r^2\right)\right.\right.\\\nonumber
&+\left.\left.3A^2r^4+8Ar^2+3\right)+2 \zeta^2r^2
\left(ABr^2+B-3A\left(Ar^2+2\right)\right)-2\right)+Br^2
\\\nonumber &\times\left(2 \zeta^3-9\zeta^2+8 \zeta+\left(2
\zeta^3-7\zeta^2+10\zeta-1\right)Ar^2-1\right)+\zeta^2-2\zeta^3A^2
r^4\\\nonumber &-2 \zeta +3 \zeta
^2A^2r^4+A^2r^4-4\zeta^3Ar^2+8\zeta^2Ar^2-8 \zeta
Ar^2+4Ar^2+1\bigg]\\\label{59a} &+4\pi
r^4\left(\mathcal{T}^{1(\mathcal{D})}_{1}
-\mathcal{T}^{2(\mathcal{D})}_{2}\right),\\\nonumber
\hat{\Pi}&=\frac{1}{8\pi}\left[\frac{e^{-Br^2}}{2r^2\left(\zeta
\left(e^{Br^2}-1\right)+1\right)^2}\left\{-e^{Br^2} \left(\left(2
B^2 r^4-10B\left(Ar^4+r^2\right)+6A^2
r^4\right.\right.\right.\right.\\\nonumber
&+\left.16Ar^2+3\right)\zeta^2+2\zeta^3r^2
\left(2B\left(Ar^2+1\right)-3A\left(Ar^2+2\right)\right)+2\zeta
\left(\left(4Ar^4\right.\right.\\\nonumber
&+\left.\left.\left.r^2\right)B-4Ar^2-2\right)+1\right)+2B^2 \zeta
r^4+\zeta^2e^{3Br^2}\left(2\zeta A^2r^4+4\zeta
Ar^2-1\right)\\\nonumber &+\zeta e^{2Br^2} \left(\zeta
\left(-B\left(3Ar^4+r^2\right)+3A^2 r^4+8Ar^2+3\right)+2\zeta^2 r^2
\left(ABr^2+B\right.\right.\\\nonumber
&-\left.\left.3A\left(Ar^2+2\right)\right)-2\right)+Br^2 \left(2
\zeta^3-9\zeta^2+8\zeta+\left(2\zeta^3-7\zeta^2+10\zeta
-1\right)\right.\\\nonumber &\times\left.Ar^2-1\right)+\zeta^2-2
\zeta -2\zeta^3A^2 r^4+3 \zeta ^2A^2 r^4+A^2 r^4-4 \zeta ^3Ar^2+8
\zeta ^2Ar^2\\\nonumber &-\left.8 \zeta
Ar^2+4Ar^2+1\right\}-\frac{e^{-Br^2}}{2}\left\{A\left(4+Ar^2\right)+B\left(4\zeta-1-Ar^2\right)+\frac{1}{r^2}\right.\\\label{g60}
&\times\left.\left.\left(1-2\zeta\right)\right\}
+\frac{1}{2r^2}\left(1-2\zeta\right)\right].
\end{align}

\subsection{Physical Analysis of the Developed Solutions}

The mass of spherically symmetric bodies can be written as
\begin{equation}\label{g63}
m(r)=4\pi\int_{0}^{H}r^2 \hat{\mu}dr.
\end{equation}
We calculate the mass of corresponding geometry \eqref{g6} by
applying numerical technique on Eq.\eqref{g63} and use an initial
condition $m(0)=0$. A self-gravitating system can be described by
its various physical properties, one of them is the compactness
parameter $\big(\sigma(r)\big)$ which presents the ratio of mass and
radius of that system. The maximum value of parameter $\sigma(r)$
was found by Buchdahl \cite{42a} by calculating the matching
conditions of corresponding inner and outer geometries at the
hypersurface. He observed that this limit should not be greater than
$\frac{4}{9}$ for the case of stable configuration. A celestial
structure having a robust gravitational pull diffuses
electromagnetic radiations due to some reactions occurring in the core of that
body. The wavelength of such radiations increases with time and this
can be computed by a redshift parameter $\big(D(r)\big)$. It is
characterized as $D(r)=\frac{1}{\sqrt{1-2\sigma}}-1$. Buchdahl
restricted its value as $D(r)<2$ for ideal stable configuration,
while it was observed to be 5.211 for the case of matter
distribution involving pressure anisotropy \cite{42b}.

Another phenomenon of great importance in astrophysics is the energy
conditions. The agreement with such constraints guarantees the
presence of usual matter as well as viable solutions. The matter
variables which represent the interior configuration of a compact
object (involving ordinary matter) must satisfy these bounds. The
energy conditions are classified into four types which take the form
in $f(\mathcal{R},\mathcal{T},\mathcal{Q})$ gravitational theory as
\begin{eqnarray}\nonumber
&&\hat{\mu}+\frac{s^2}{8\pi r^4} \geq 0, \quad \hat{\mu}+\hat{p}_{r}
\geq 0,\\\nonumber &&\hat{\mu}+\hat{p}_{\bot}+\frac{s^2}{4\pi r^4}
\geq 0, \quad \hat{\mu}-\hat{p}_{r}+\frac{s^2}{4\pi r^4} \geq
0,\\\label{g50} &&\hat{\mu}-\hat{p}_{\bot} \geq 0, \quad
\hat{\mu}+\hat{p}_{r}+2\hat{p}_{\bot}+\frac{s^2}{4\pi r^4} \geq 0.
\end{eqnarray}
The stability of cosmological solutions plays a crucial role in the
field of astrophysics to check their feasibility. In this regard, we
utilize two different approaches to investigate the regions in inner
spacetime where both of the obtained solutions are stable. Firstly,
we employ causality condition \cite{42c} which declares that the
squared sound speed should be within $(0,1)$, i.e., $0 <
v_{s}^{2} < 1$. The Herrera's cracking approach states that absolute
value of the difference between squared sound speeds in both
tangential $(v_{s\bot}^{2}=\frac{d\hat{p}_{\bot}}{d\hat{\mu}})$ and
radial directions $(v_{sr}^{2}=\frac{d\hat{p}_{r}}{d\hat{\mu}})$
should be less than $1$ for the case of stable anisotropic
configuration \cite{42ca}. Mathematically, the compact object is
stable if $0 < \mid v_{s\bot}^{2}-v_{sr}^{2} \mid < 1$ holds. Another
key factor which is used to determine the stability of compact
geometry is the adiabatic index $(\Lambda)$. An astronomical object
is stable in the domain where the index $(\Lambda)$ gains its value
greater than $\frac{4}{3}$ \cite{42d}-\cite{42f}. For this gravity,
$\Lambda$ is defined as
\begin{equation}\label{g62}
\hat{\Lambda}=\frac{\hat{\mu}+\hat{p}_{r}}{\hat{p}_{r}}
\left(\frac{d\hat{p}_{r}}{d\hat{\mu}}\right).
\end{equation}
\begin{figure}\center
\epsfig{file=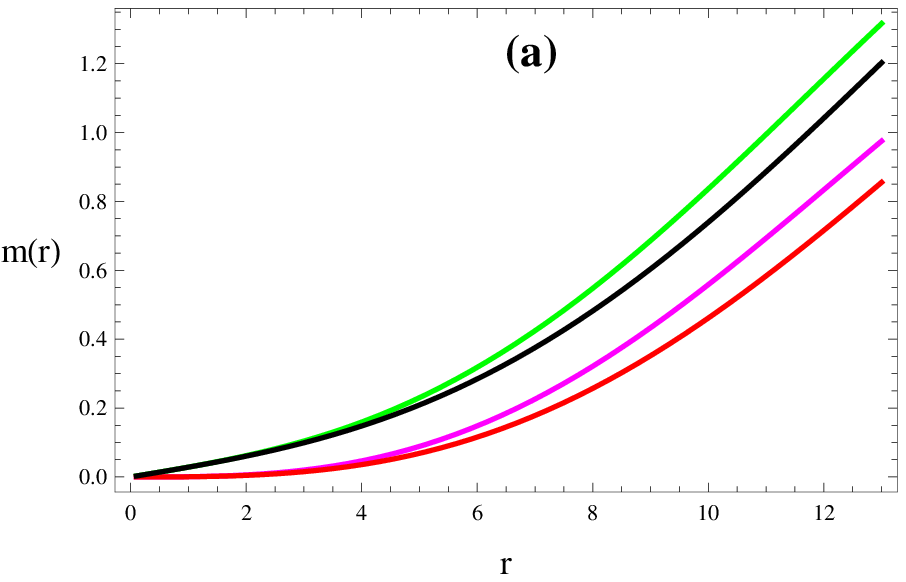,width=0.4\linewidth}
\epsfig{file=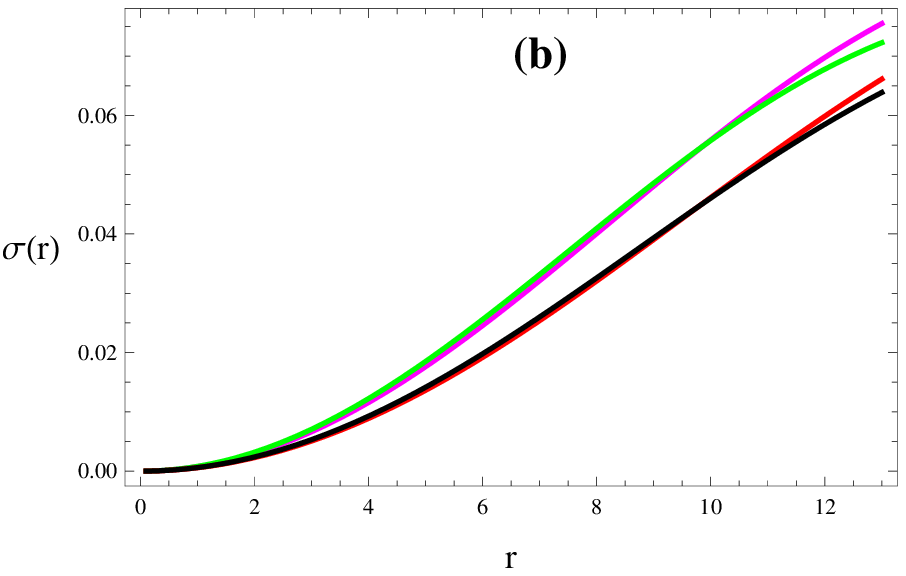,width=0.4\linewidth}
\epsfig{file=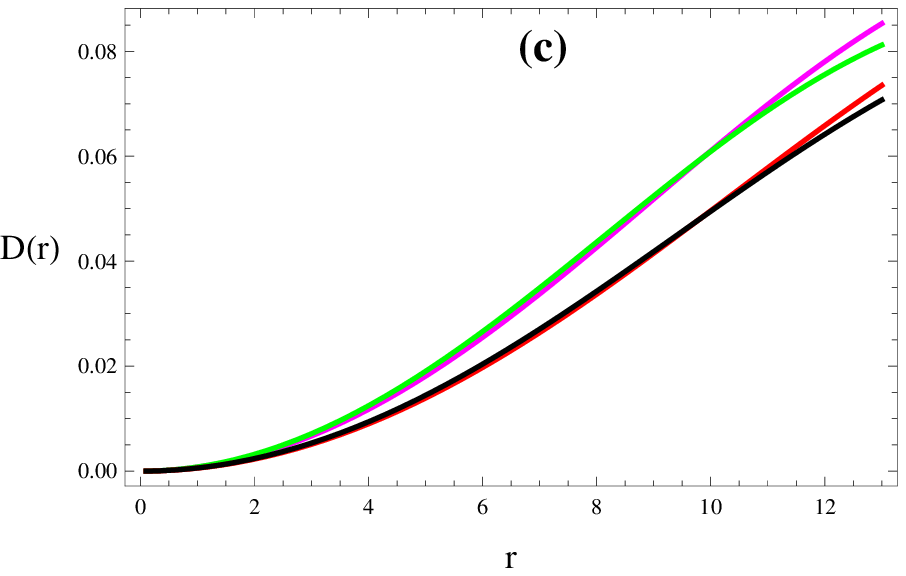,width=0.4\linewidth} \caption{Plots of
mass (in km) (\textbf{a}), compactness (\textbf{b}) and redshift
(\textbf{c}) parameters corresponding to
$\mathcal{S}=0.1,~\zeta=0.5$ (pink),~$\zeta=0.9$ (green) and
$\mathcal{S}=0.8,~\zeta=0.5$ (red),~$\zeta=0.9$ (black) for
solution-I}
\end{figure}

The $f(\mathcal{R},\mathcal{T},\mathcal{Q})$ theory comprises the
complicated equations of motion due to the factor
$\mathcal{Q}=\mathcal{R}_{\lambda\xi}\mathcal{T}^{\lambda\xi}$.
Therefore for our convenience, we choose a linear model \cite{22} to
explore physical features of the developed solutions by taking
arbitrary values of constant $\varrho$ as
\begin{equation}\label{g61}
f(\mathcal{R},\mathcal{T},\mathcal{R}_{\lambda\xi}\mathcal{T}^{\lambda\xi})=\mathcal{R}+\varrho
\mathcal{R}_{\lambda\xi}\mathcal{T}^{\lambda\xi}.
\end{equation}
The contraction of energy-momentum tensor with the Ricci tensor in
the above model ensures that massive test particles in the
gravitational field of self-gravitating model still entails the
effects of non-minimal matter-geometry interaction. Here, the value
of $\varrho$ can be negative or positive. The positive values of
this arbitrary constant provide unacceptable behavior of the matter
variables such as energy density and radial/tangential pressures
corresponding to both the obtained solutions, as their values appear
in negative range. Consequently, the solutions are no more viable as
well as stable. Thus, we have the only choice for its negative
values. First, we check the physical behavior of solution-I for
$\varrho=-0.1$ and the constant $B$ defined in Eq.\eqref{g43}. The
other two constants $A$ and $C$ are shown in Eqs.\eqref{g37} and
\eqref{g38a}. We plot the graphs for mass, compactness and redshift
of compact sphere \eqref{g6} corresponding to the decoupling
parameter $\zeta=0.5$ and $0.9$ in Figure \textbf{1}. The mass shows
increasing behavior with rise in $\zeta$ while charge decreases its
value linearly. The particular values of $\zeta$ as well as charge
confirm the compactness and redshift factors within their required
limits, as shown in Figure \textbf{1} (\textbf{b},\textbf{c}).
\begin{figure}\center
\epsfig{file=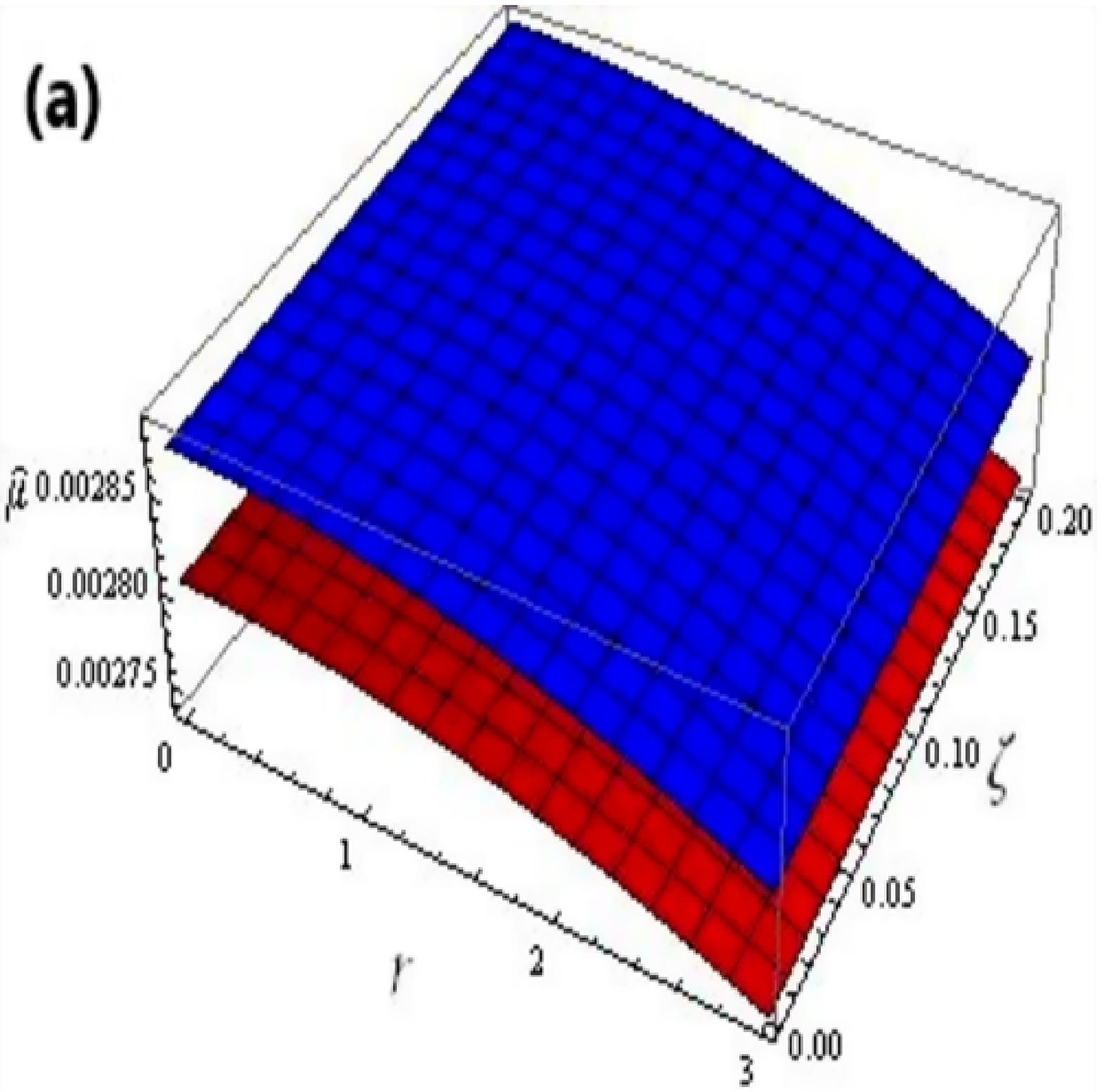,width=0.4\linewidth}\epsfig{file=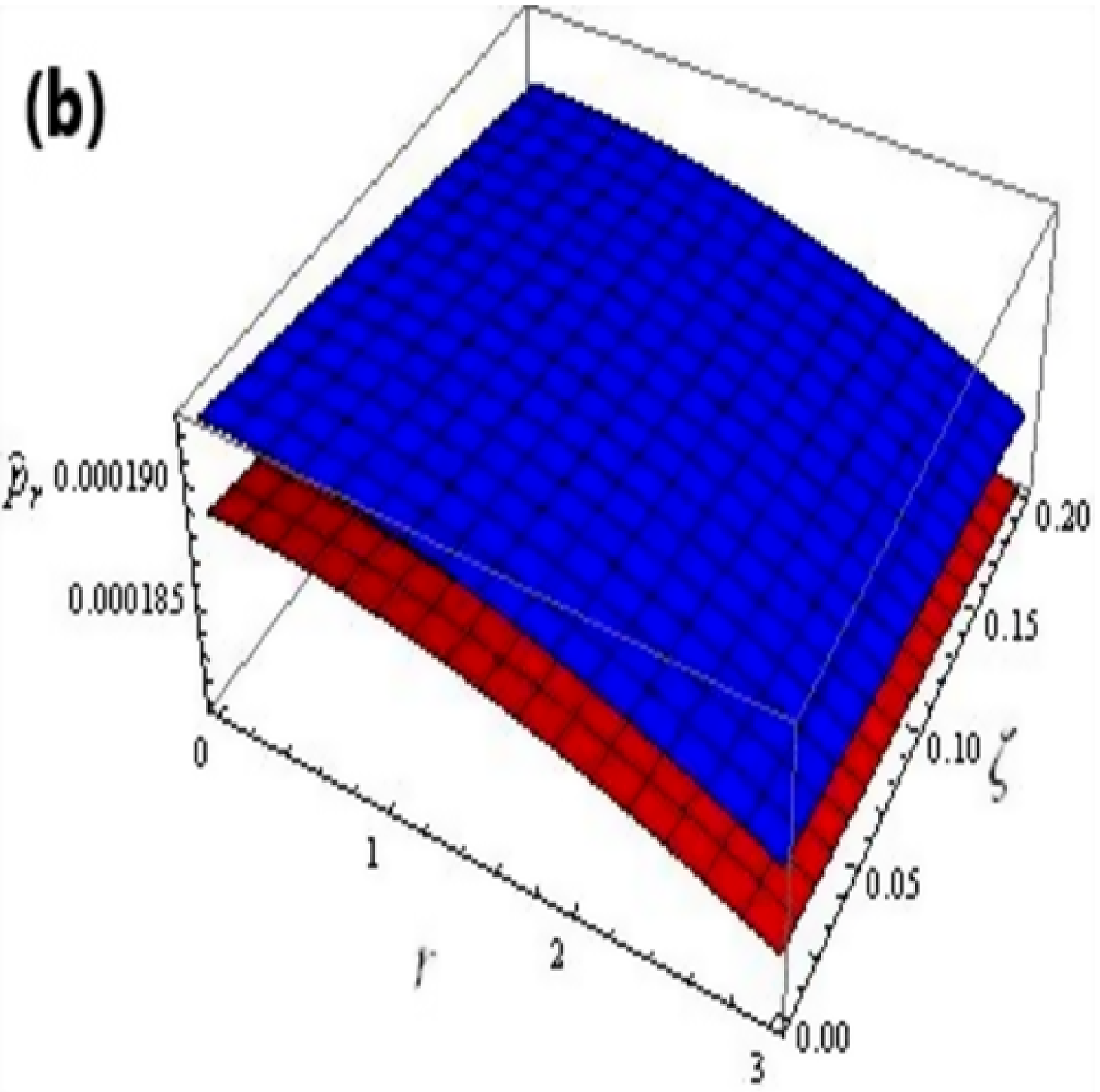,width=0.4\linewidth}
\epsfig{file=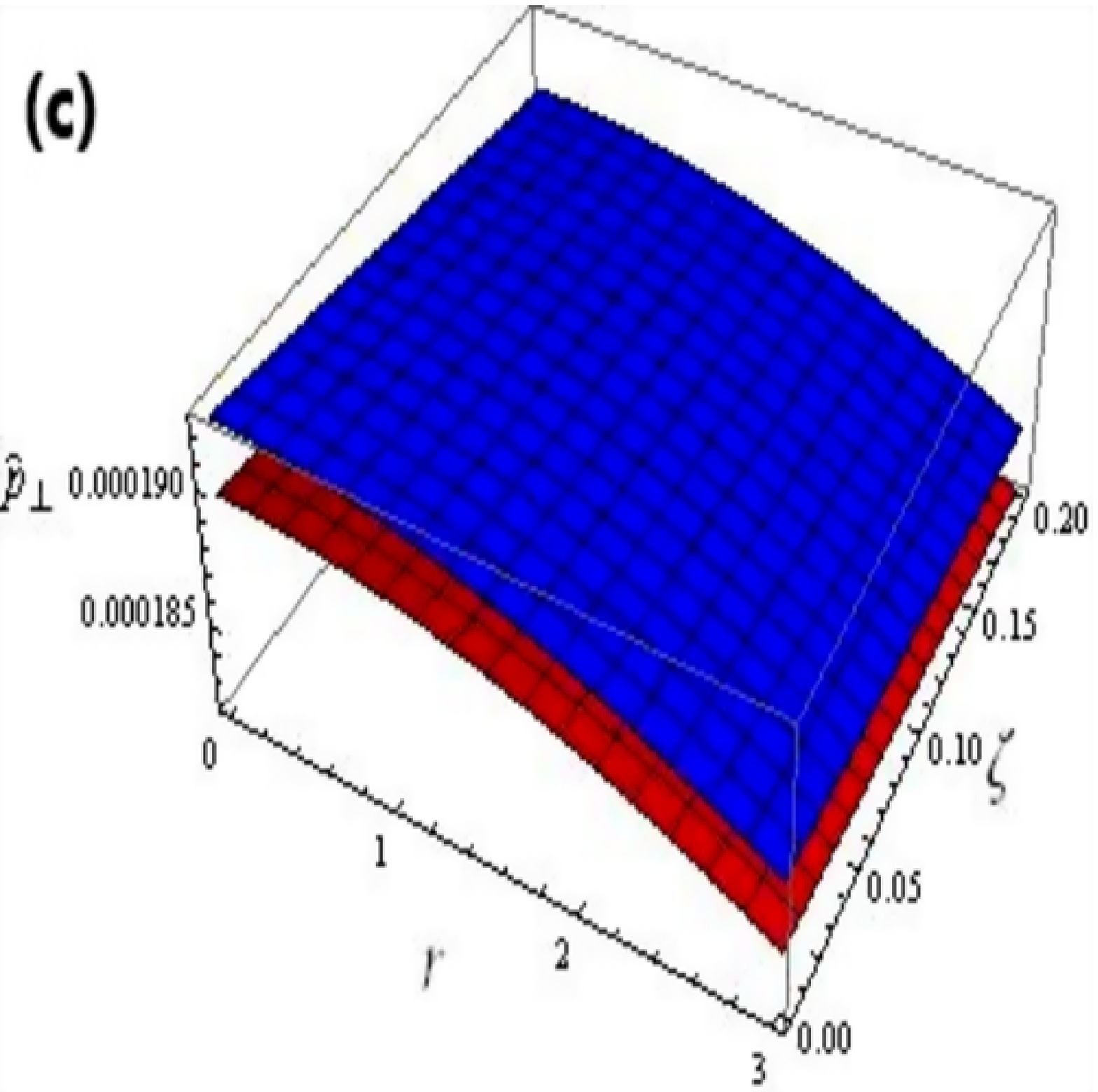,width=0.4\linewidth}\epsfig{file=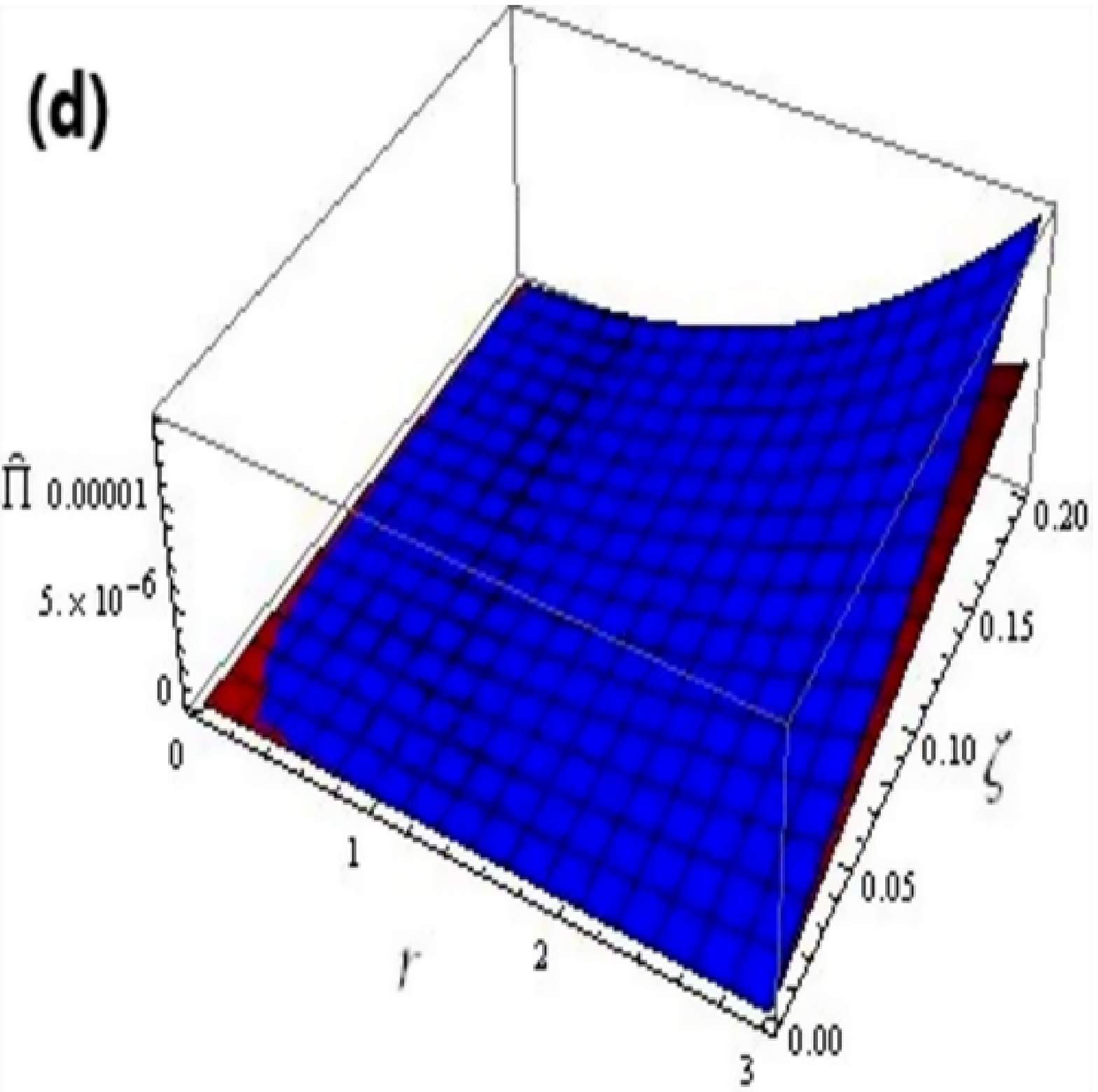,width=0.4\linewidth}
\caption{Plots of energy density (in km$^{-2}$) (\textbf{a}), radial
pressure (in km$^{-2}$) (\textbf{b}), tangential pressure (in
km$^{-2}$) (\textbf{c}) and anisotropy (in km$^{-2}$) (\textbf{d})
versus $r$ and $\zeta$ with $\mathcal{S}=0.1$ (Blue),
$\mathcal{S}=0.8$ (Red), $\mathcal{M}=1M_{\bigodot}$ and
$H=(0.2)^{-1}M_{\bigodot}$ for solution-I}
\end{figure}
\begin{figure}\center
\epsfig{file=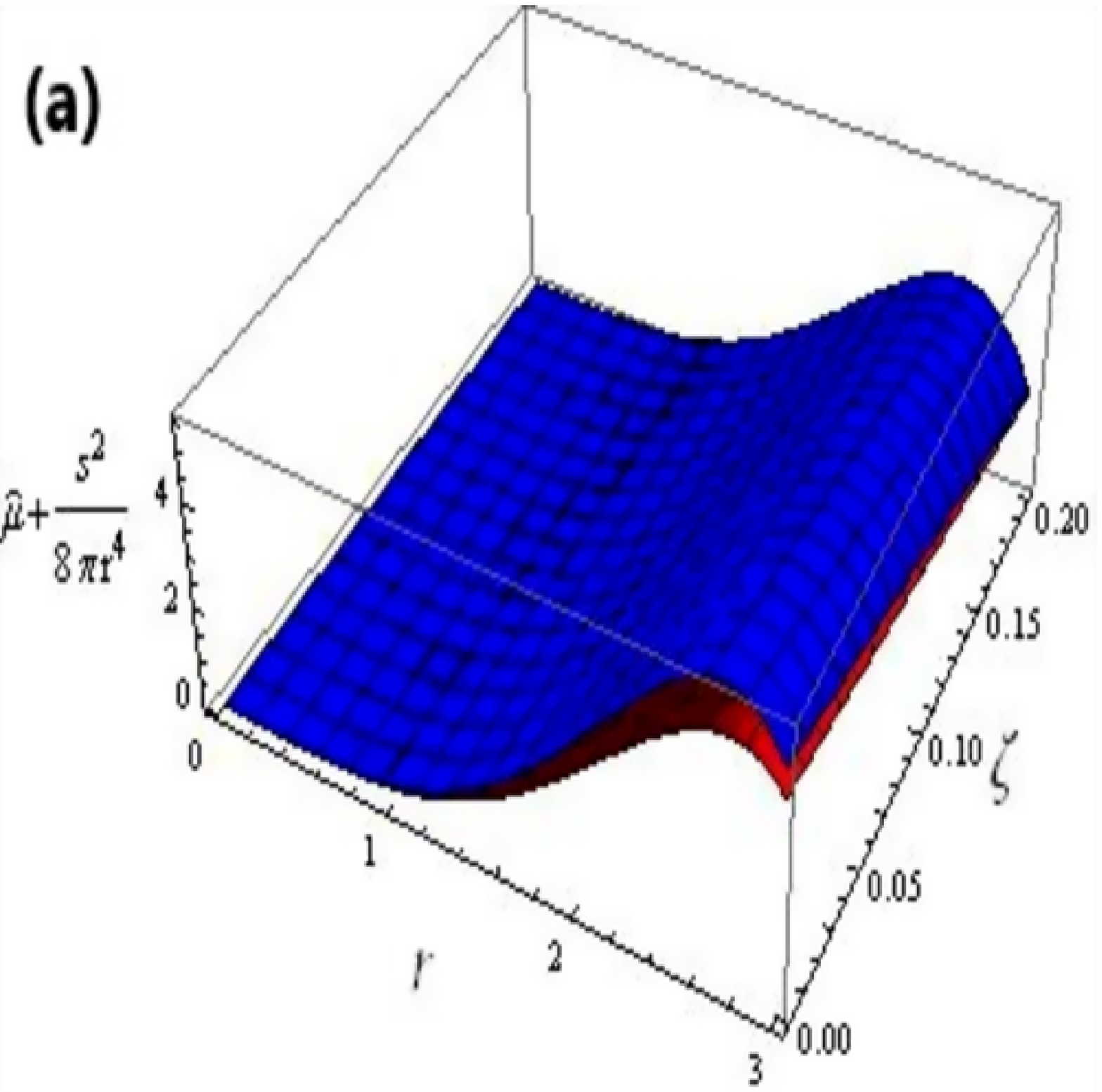,width=0.4\linewidth}\epsfig{file=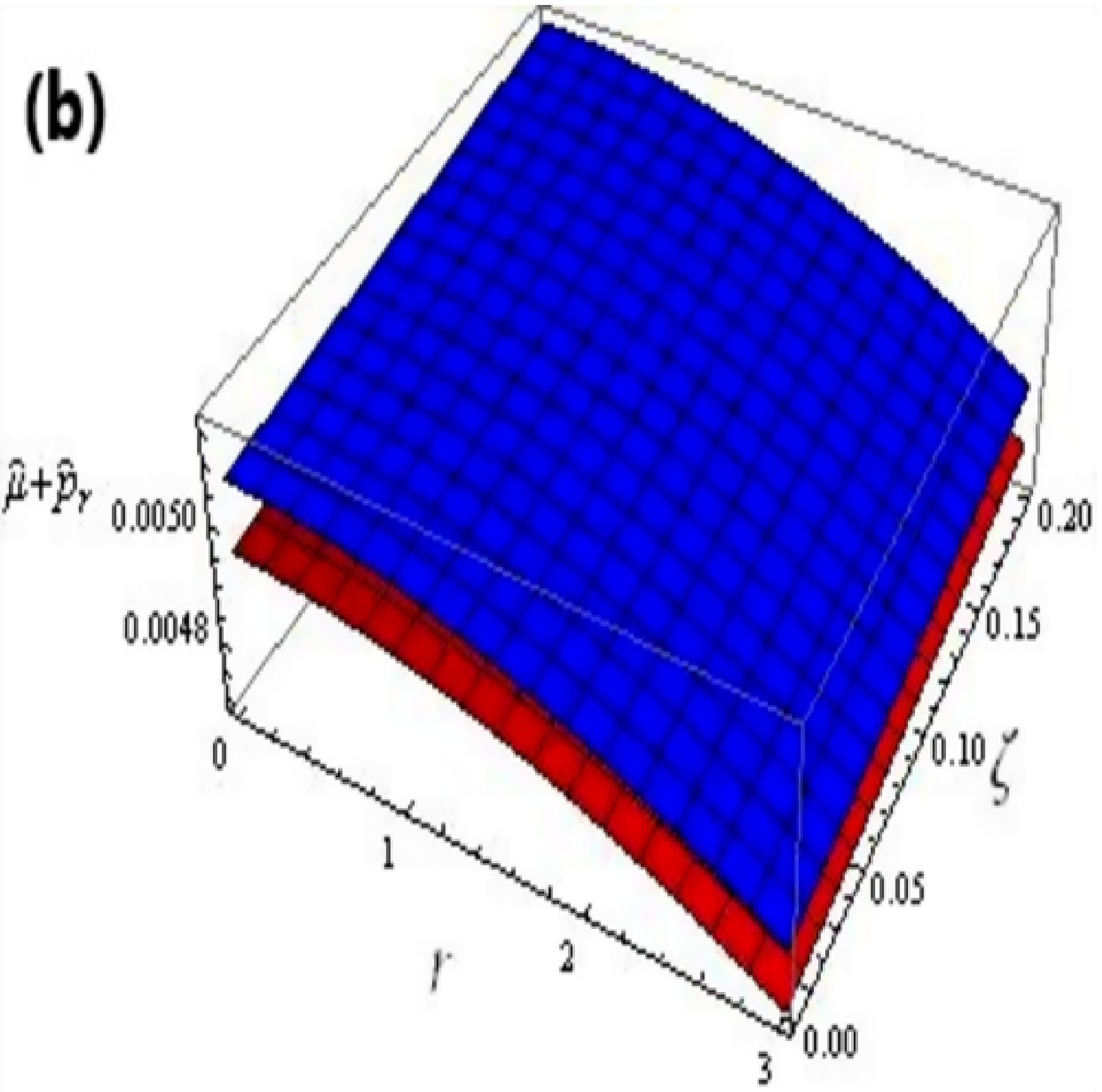,width=0.4\linewidth}
\epsfig{file=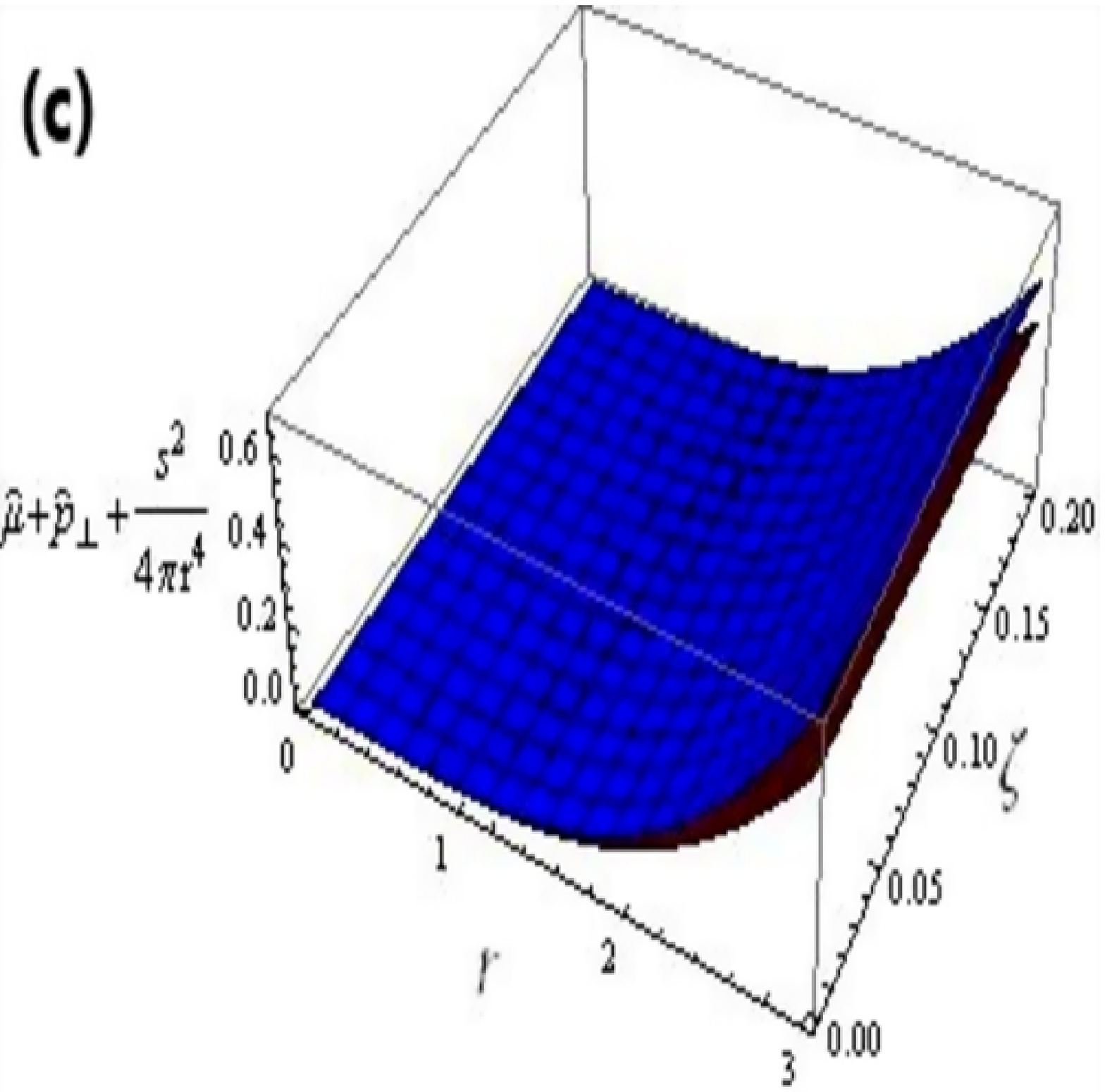,width=0.4\linewidth}\epsfig{file=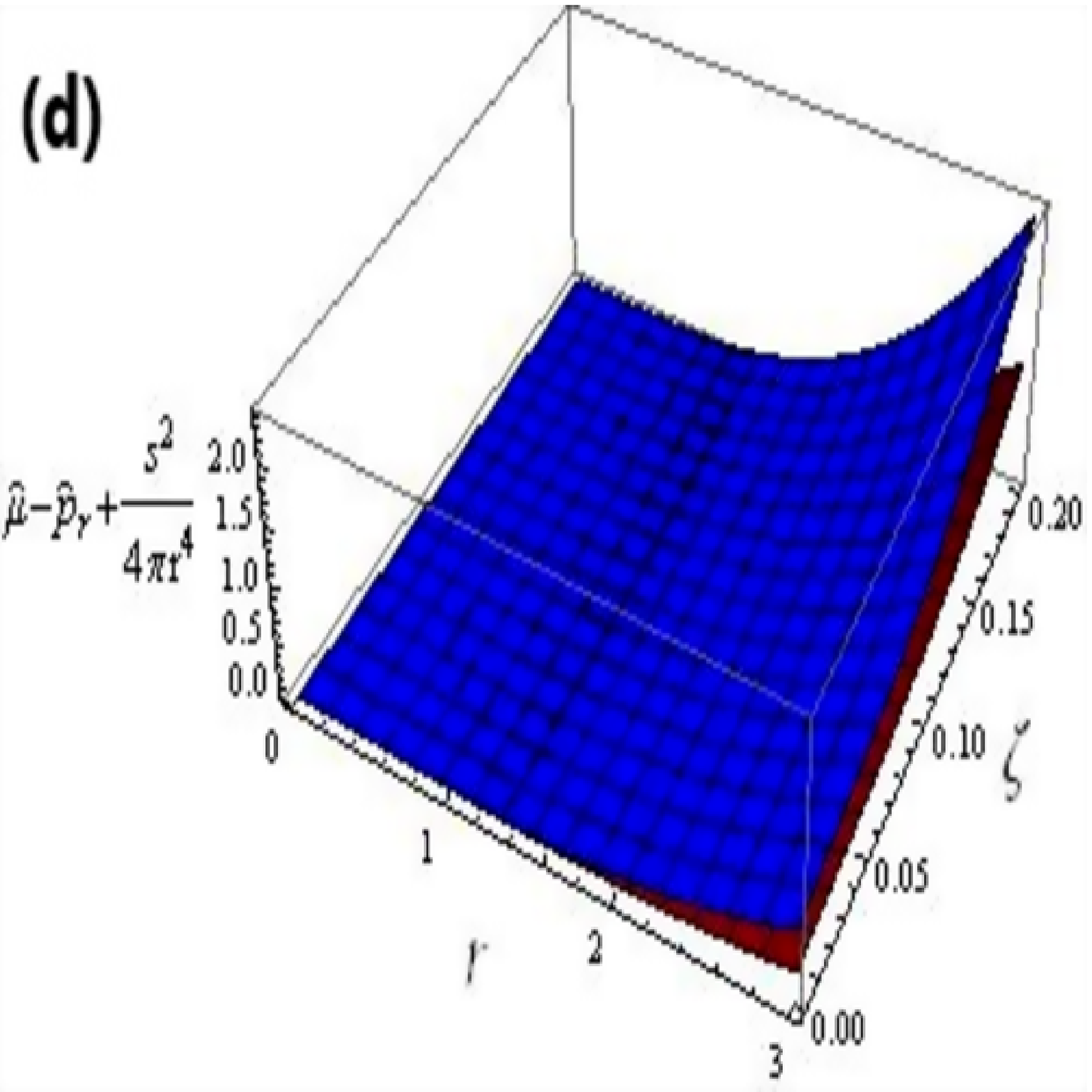,width=0.4\linewidth}
\epsfig{file=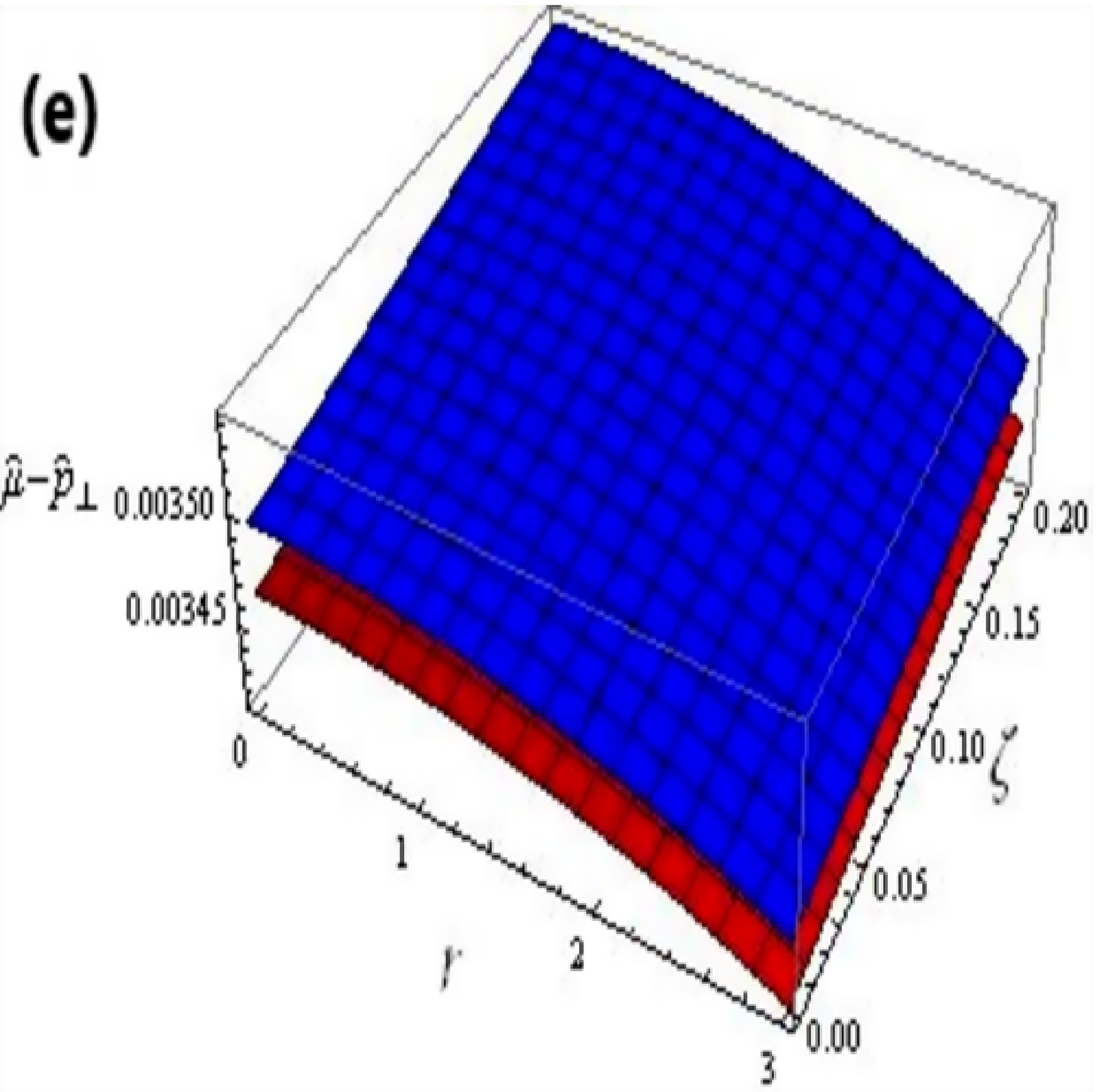,width=0.4\linewidth}\epsfig{file=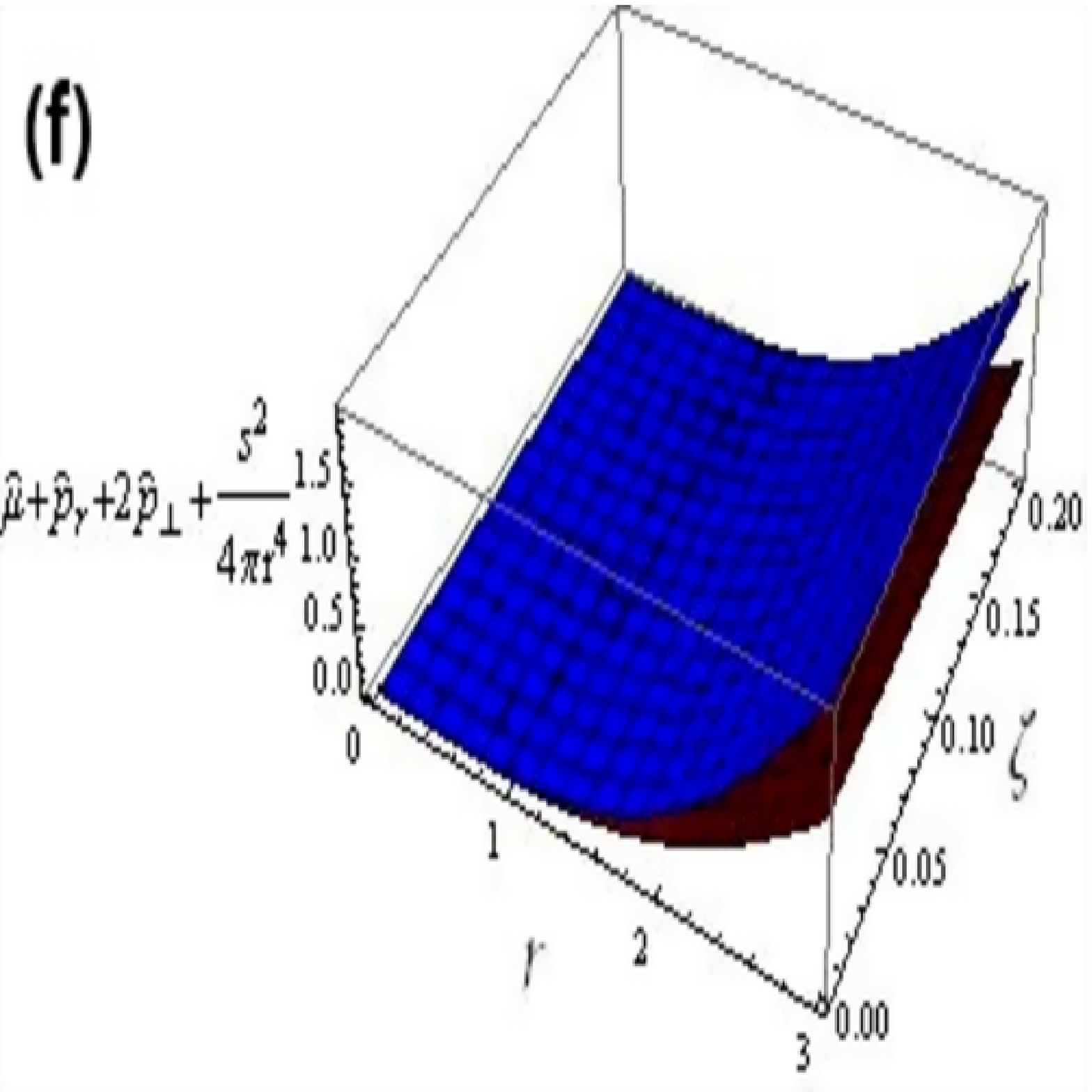,width=0.4\linewidth}
\caption{Plots of energy conditions (in km$^{-2}$) versus $r$ and
$\zeta$ with $\mathcal{S}=0.1$ (Blue), $\mathcal{S}=0.8$ (Red),
$\mathcal{M}=1M_{\bigodot}$ and $H=(0.2)^{-1}M_{\bigodot}$ for
solution-I (\textbf{a}$-$\textbf{f})}
\end{figure}

The values of material variables (pressure and energy density) for
feasible structures should be maximum, positive and finite at the
center while they show decreasing behavior towards the boundary of a
star. Figure \textbf{2} (\textbf{a}) indicates the maximum value of
energy density in the middle, whereas it shows decreasing behavior
with the increment in $r$ as well as charge. Also, the behavior of
effective energy density is monotonically rising as the decoupling
parameter enhances which represents the more dense star for larger
values of $\zeta$. Figure \textbf{2} displays that the plots of
radial and tangential pressures show similar pattern for the
parameter $\varrho$. Both graphs demonstrate the decrement with rise
in all factors such as $r$, $\zeta$ and charge. The anisotropy
$\hat{\Pi}$ disappears throughout the region for the decoupling
parameter $\zeta=0$ and enhances as $\zeta$ increases which confirms
that the additional source produces stronger anisotropy in the
system. Evaluating the fundamental features of a self-gravitating
star graphically by choosing different values of the coupling
constant $\varrho$, we deduce that very small negative values of
$\varrho$ provide the suitable behavior of physical variables. All
energy conditions \eqref{g50} corresponding to solution-I are
satisfied, hence it is physically viable as illustrated in Figure
\textbf{3}. Figure \textbf{4} guarantees the stability of solution-I
for different considered values of charge and the decoupling
parameter. From Figure \textbf{4} (\textbf{a}), we find that the
system becomes less stable with increment in charge near the
boundary.
\begin{figure}\center
\epsfig{file=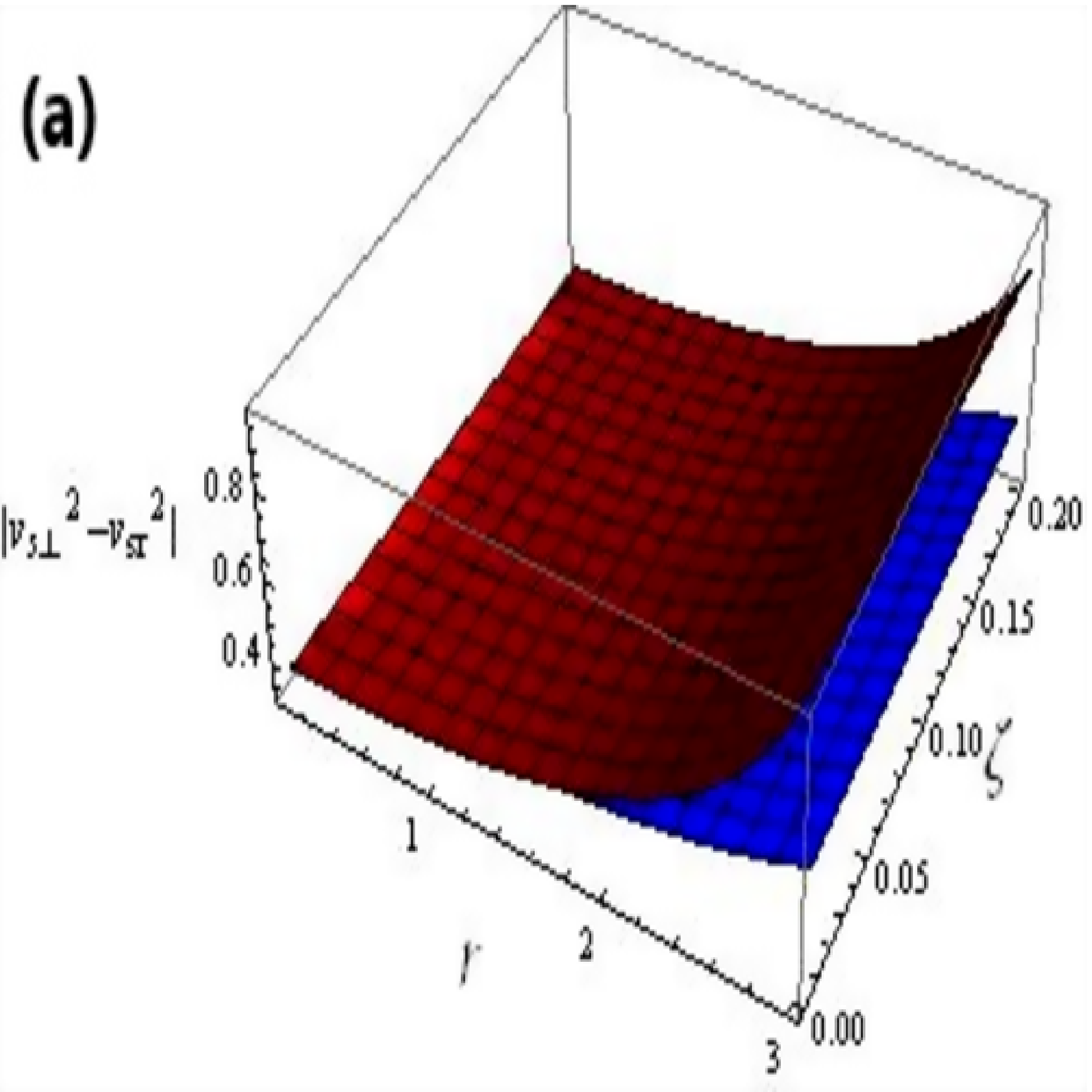,width=0.4\linewidth}\epsfig{file=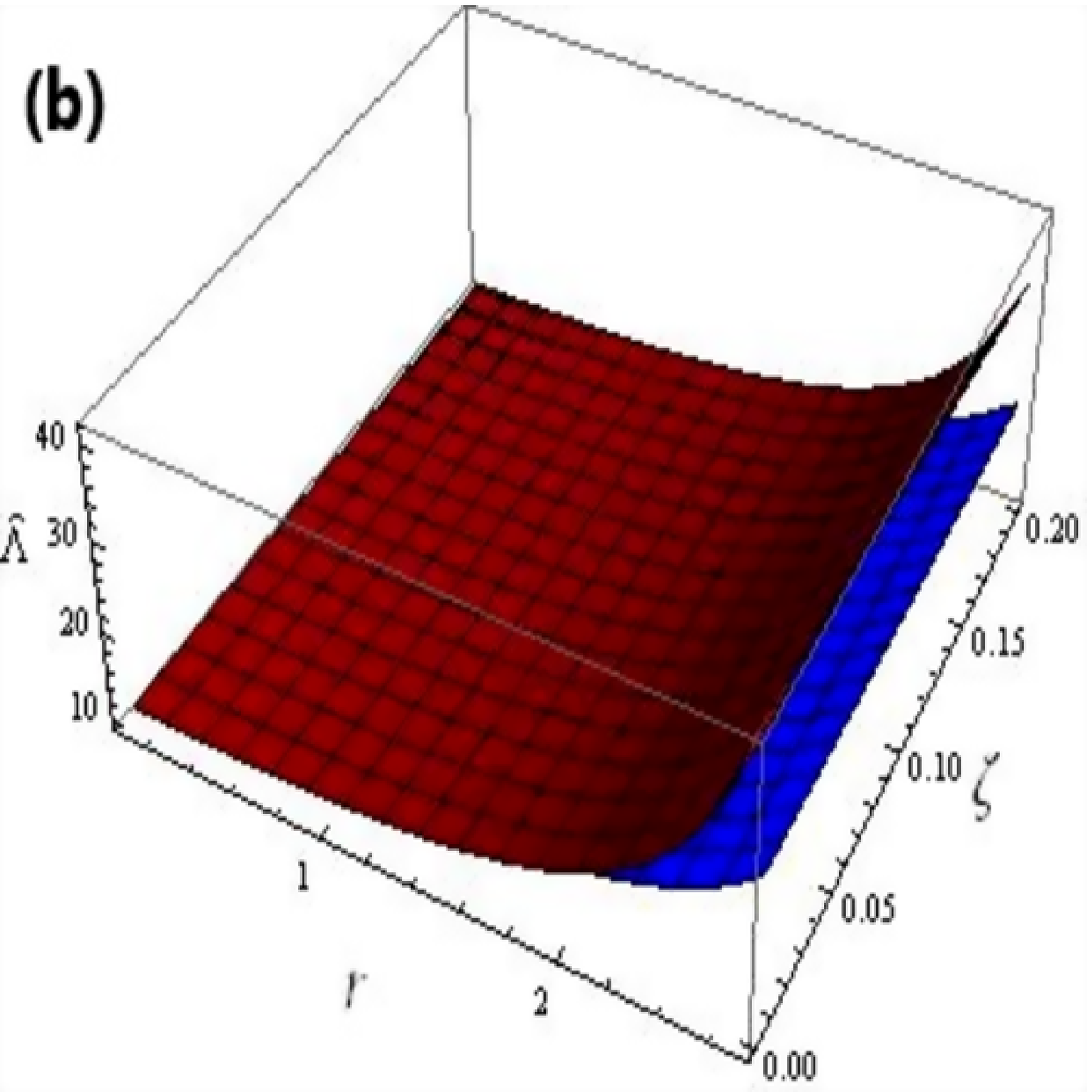,width=0.4\linewidth}
\caption{Plots of $|v_{s\bot}^2-v_{sr}^2|$ (\textbf{a}) and
adiabatic index (\textbf{b}) versus $r$ and $\zeta$ with
$\mathcal{S}=0.1$ (Blue), $\mathcal{S}=0.8$ (Red),
$\mathcal{M}=1M_{\bigodot}$ and $H=(0.2)^{-1}M_{\bigodot}$ for
solution-I}
\end{figure}

We now examine the feasibility of the obtained solution-II for
$\varrho=-0.05$. Equations \eqref{g37} and \eqref{g56} depict the
constants $A$ and $B$. We analyze the mass of geometry \eqref{g6}
for two values of the decoupling parameter $\zeta=0.1$ and
$\zeta=0.25$, as given in Figure \textbf{5} (\textbf{a}). It is
found that the mass increases with increasing $\zeta$, while the
higher value of charge yields decreasing behavior. The same figure
(\textbf{b},\textbf{c}) also shows that the compactness
$(\sigma(r))$ and redshift $(D(r))$ meet their required criteria for
both values of charge. Figure \textbf{6} illustrates the physical
behavior of different substantial variables as well as anisotropic
factor. The effective energy density and both components of
effective pressure show the same behavior as for the solution-I for
particular values of charge and $\zeta$. In the absence of $\zeta$,
anisotropy does not appear in the whole domain, while it increases
with increase in $\zeta$, as shown in Figure \textbf{6}
(\textbf{d}). Figure \textbf{7} guarantees the viability of our
second solution as all energy conditions \eqref{g50} are fulfilled.
Figure \textbf{8} confirms the stability of solution-II for
particular values of the parameter $\zeta$. It is noted from Figure
\textbf{8} (\textbf{a}) that increment in charge leads to the less
stable system for larger values of $\zeta$ near the boundary.
\begin{figure}\center
\epsfig{file=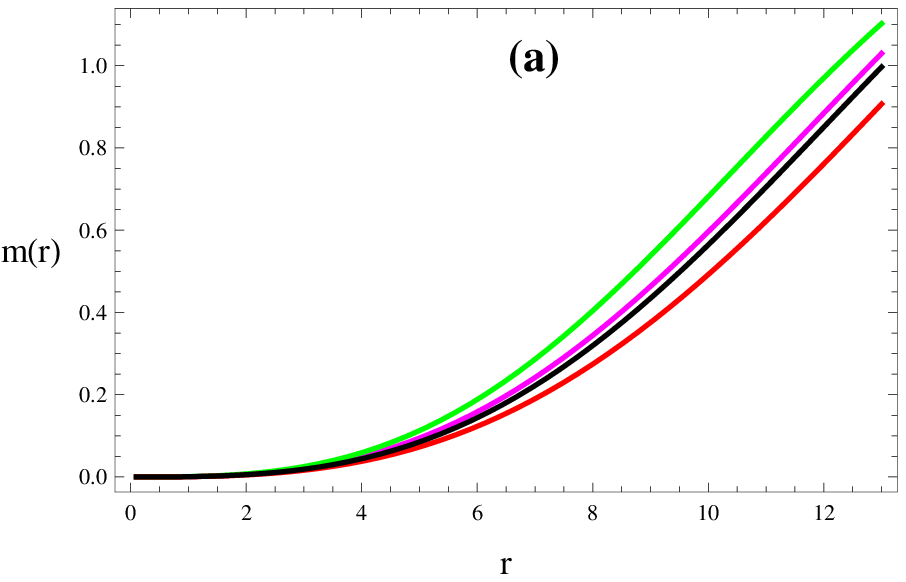,width=0.4\linewidth}
\epsfig{file=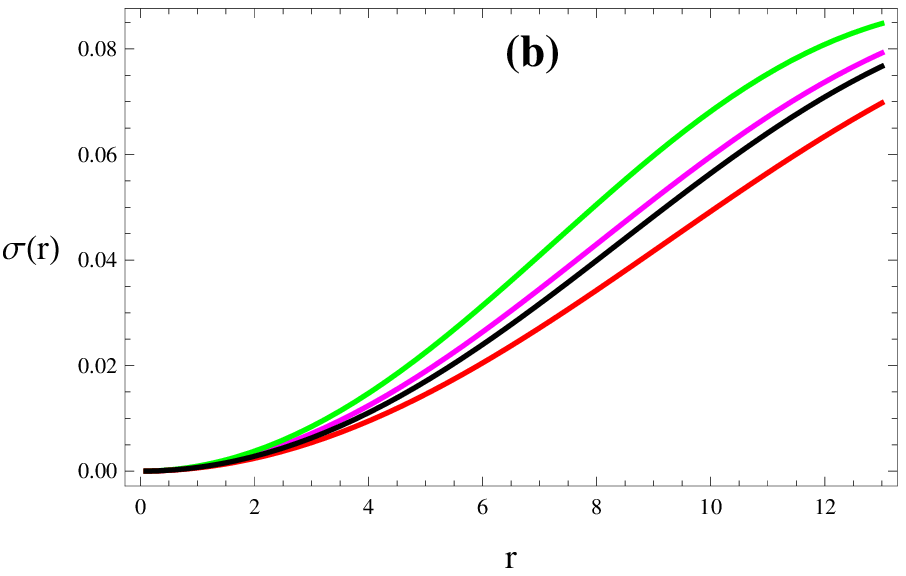,width=0.4\linewidth}
\epsfig{file=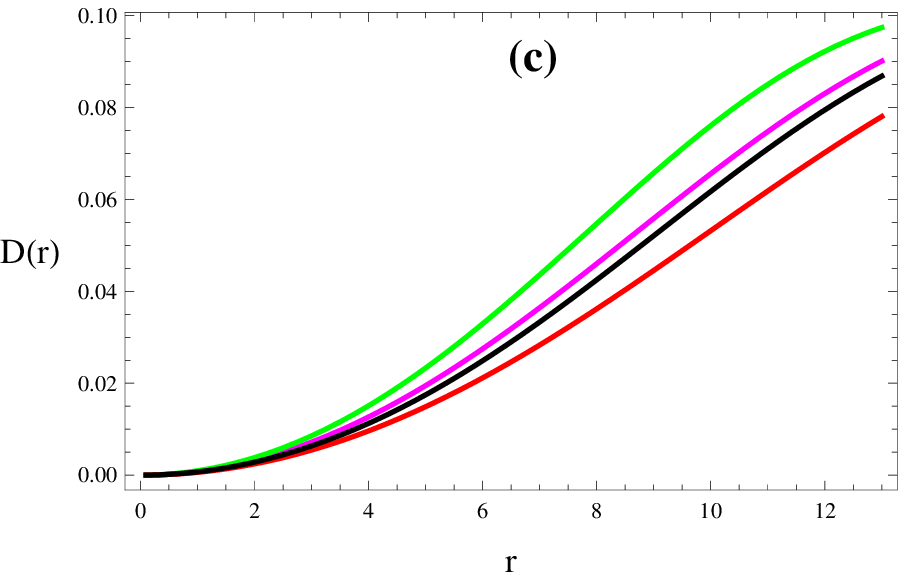,width=0.4\linewidth} \caption{Plots of
mass (in km) (\textbf{a}), compactness (\textbf{b}) and redshift
(\textbf{c}) parameters corresponding to
$\mathcal{S}=0.1,~\zeta=0.1$ (pink),~$\zeta=0.25$ (green) and
$\mathcal{S}=0.8,~\zeta=0.1$ (red),~$\zeta=0.25$ (black) for
solution-II}
\end{figure}
\begin{figure}\center
\epsfig{file=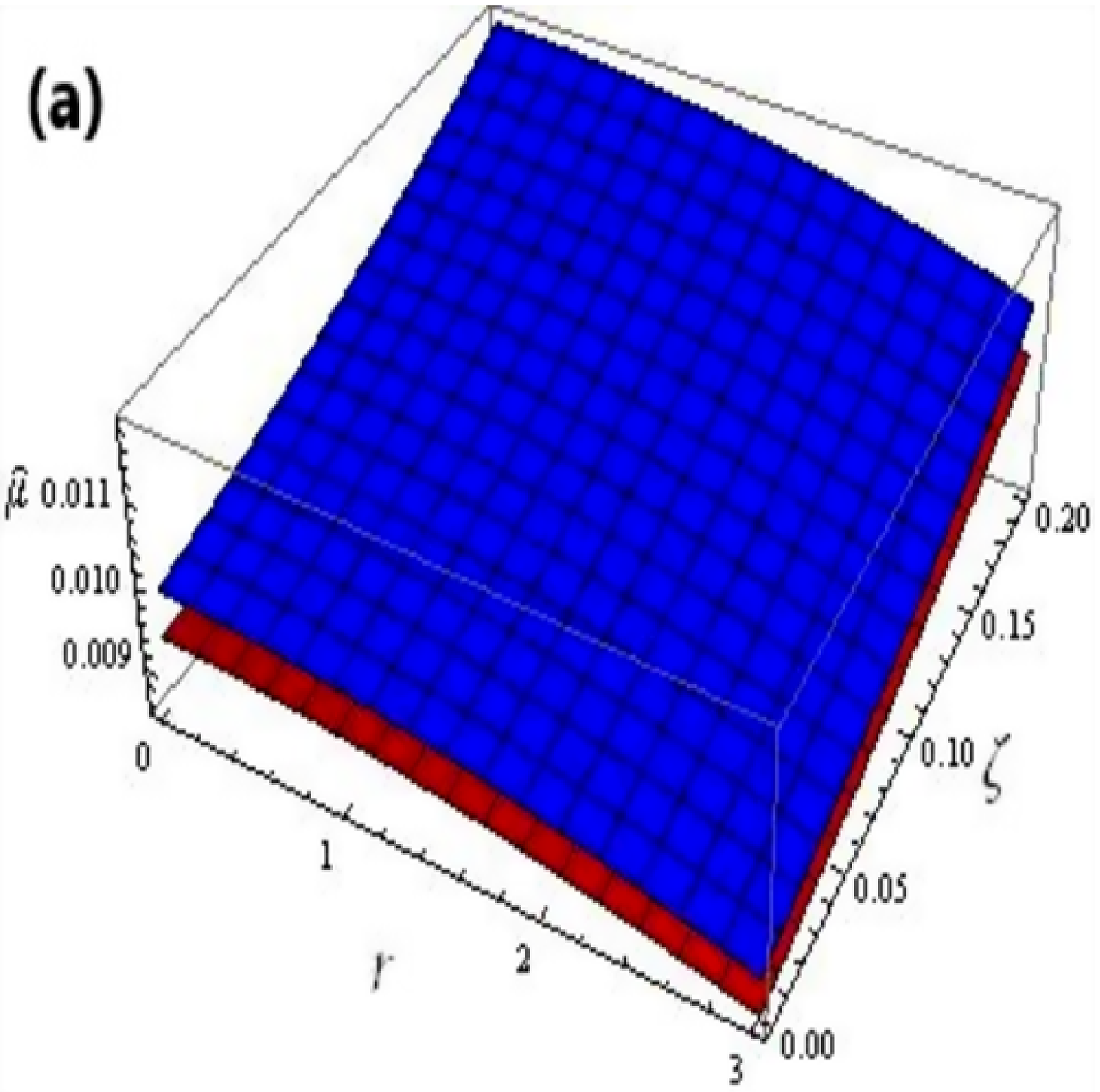,width=0.4\linewidth}\epsfig{file=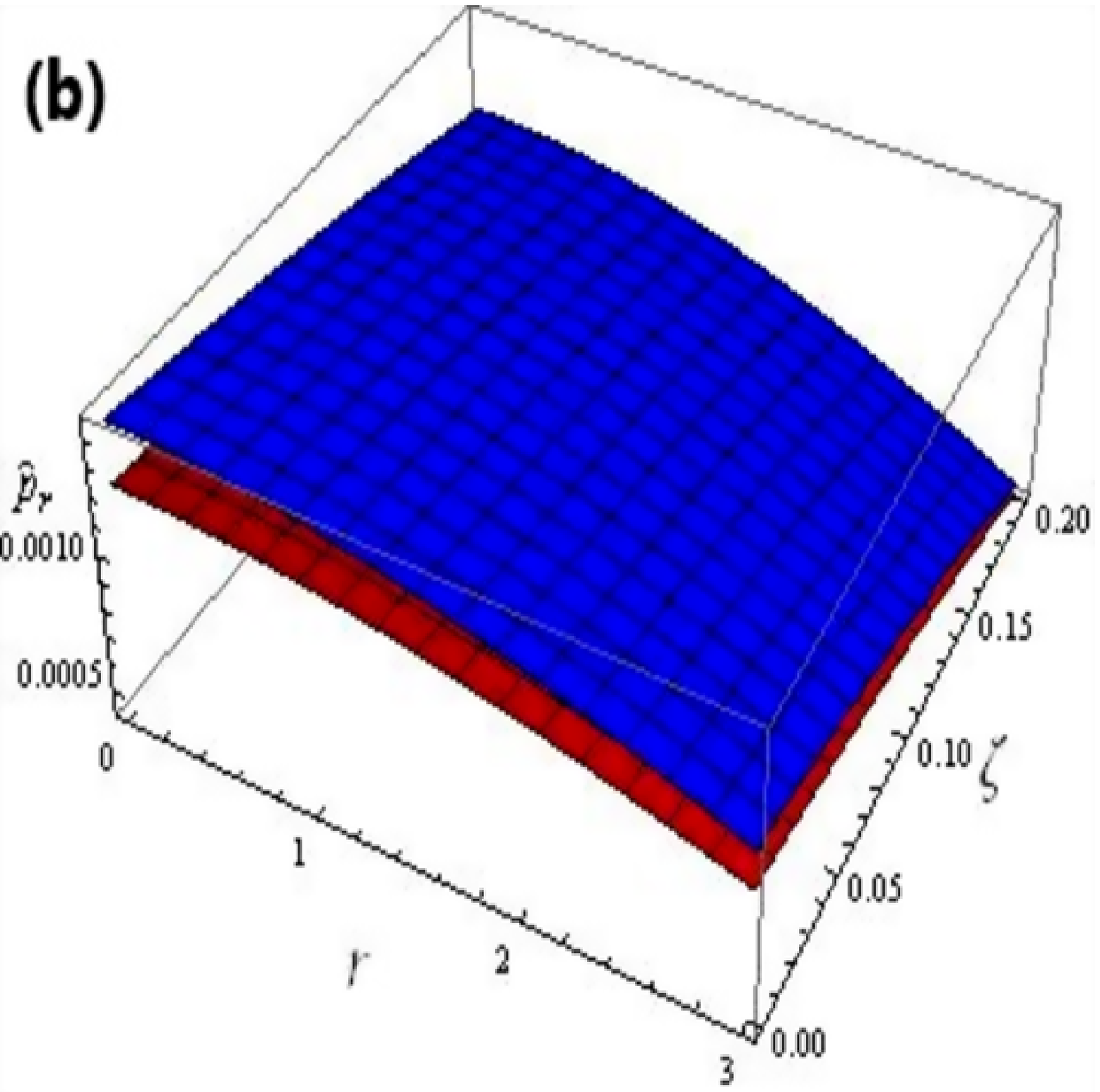,width=0.4\linewidth}
\epsfig{file=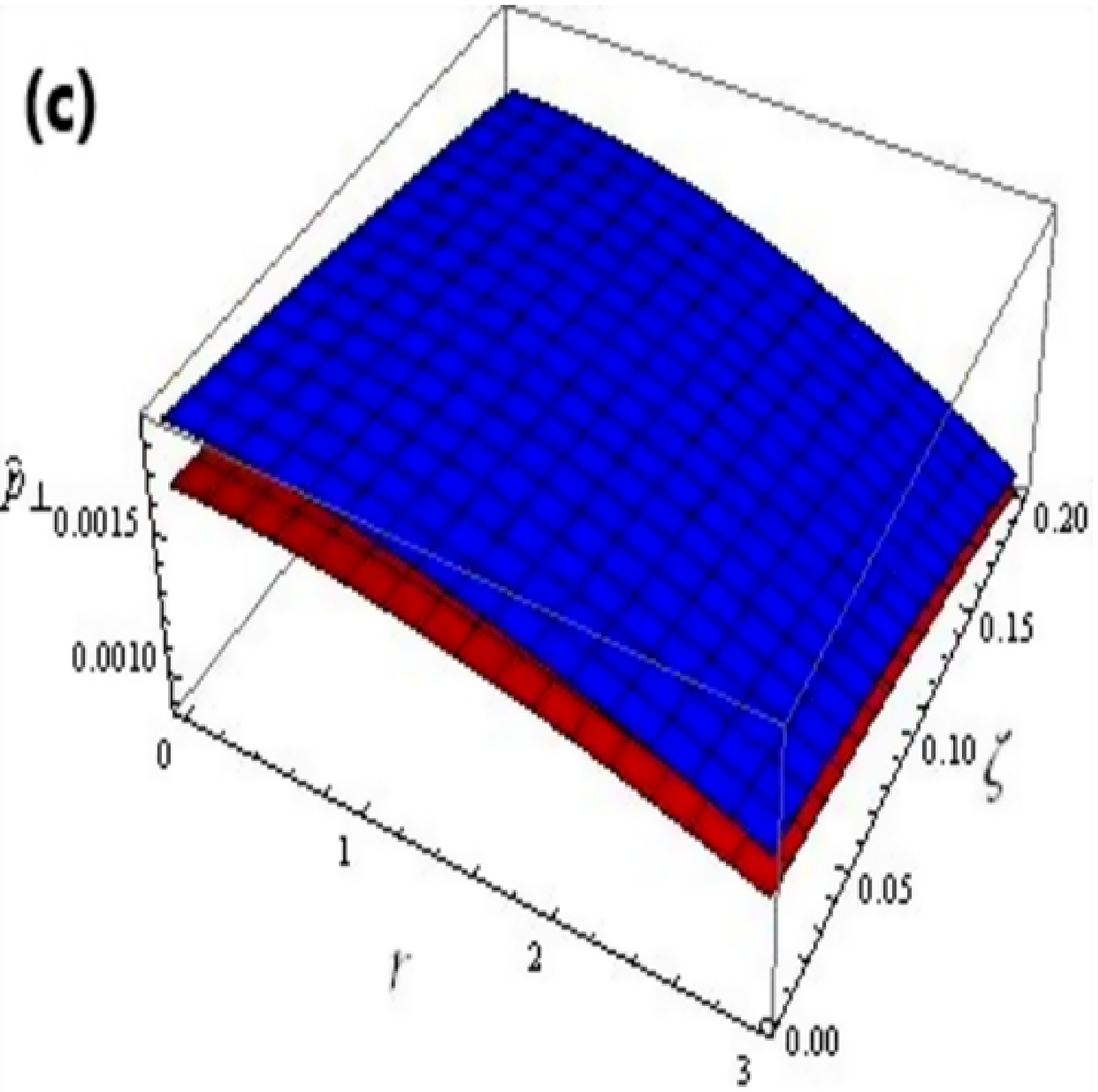,width=0.4\linewidth}\epsfig{file=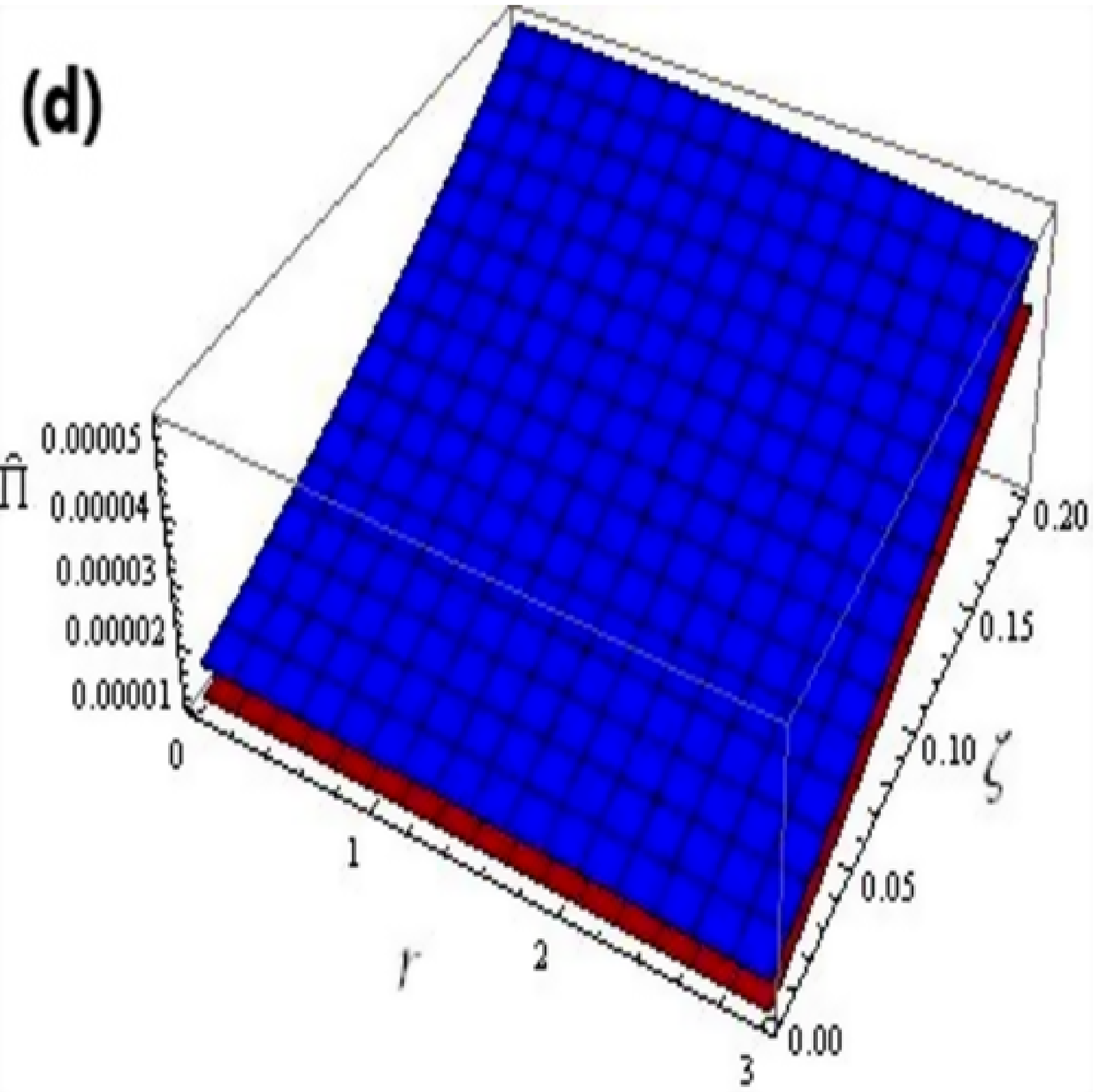,width=0.4\linewidth}
\caption{Plots of energy density (in km$^{-2}$) (\textbf{a}), radial
pressure (in km$^{-2}$) (\textbf{b}), tangential pressure (in
km$^{-2}$) (\textbf{c}) and anisotropy (in km$^{-2}$) (\textbf{d})
versus $r$ and $\zeta$ with $\mathcal{S}=0.1$ (Blue),
$\mathcal{S}=0.8$ (Red), $\mathcal{M}=1M_{\bigodot}$ and
$H=(0.2)^{-1}M_{\bigodot}$ for solution-II}
\end{figure}
\begin{figure}\center
\epsfig{file=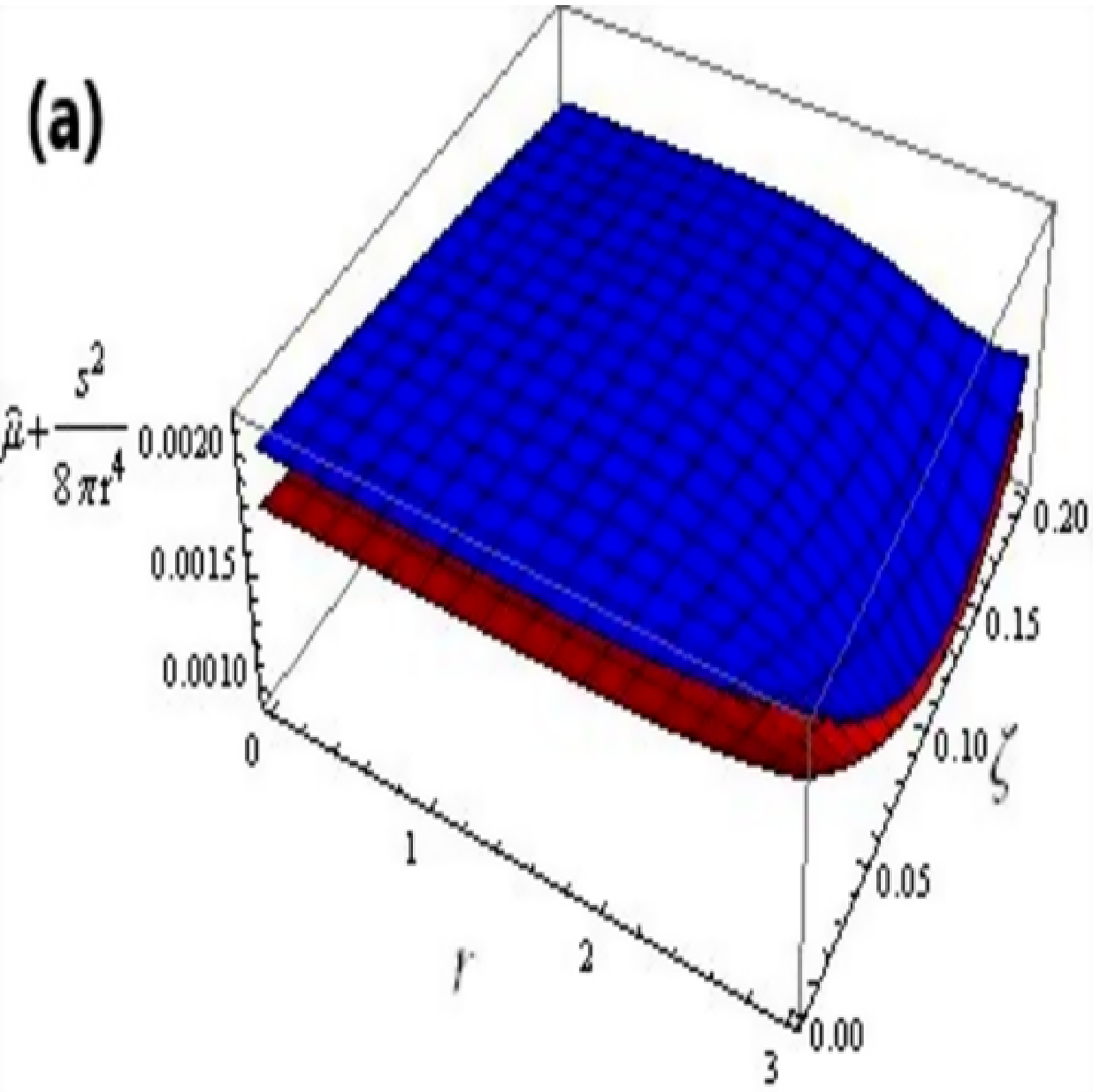,width=0.4\linewidth}\epsfig{file=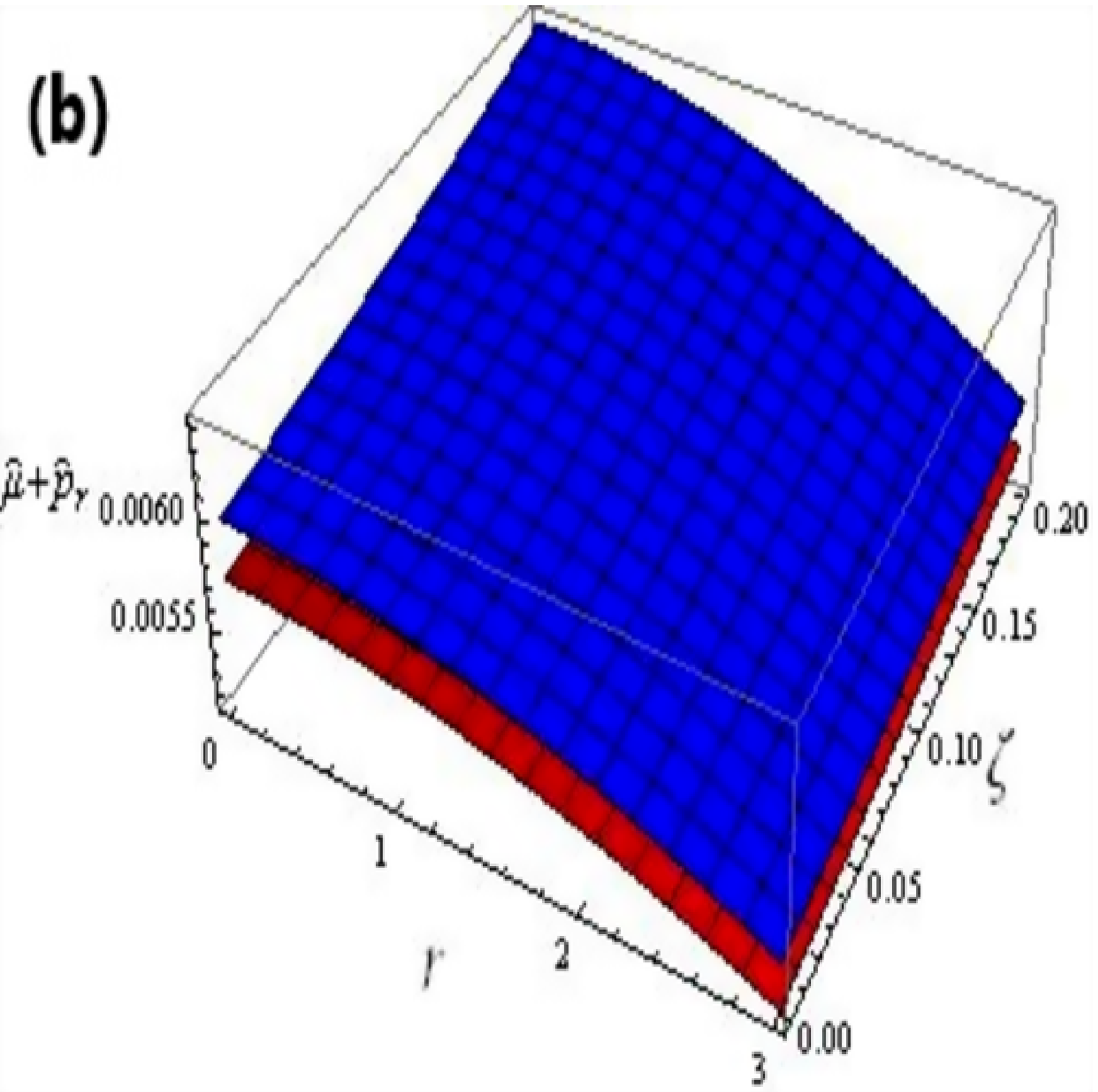,width=0.4\linewidth}
\epsfig{file=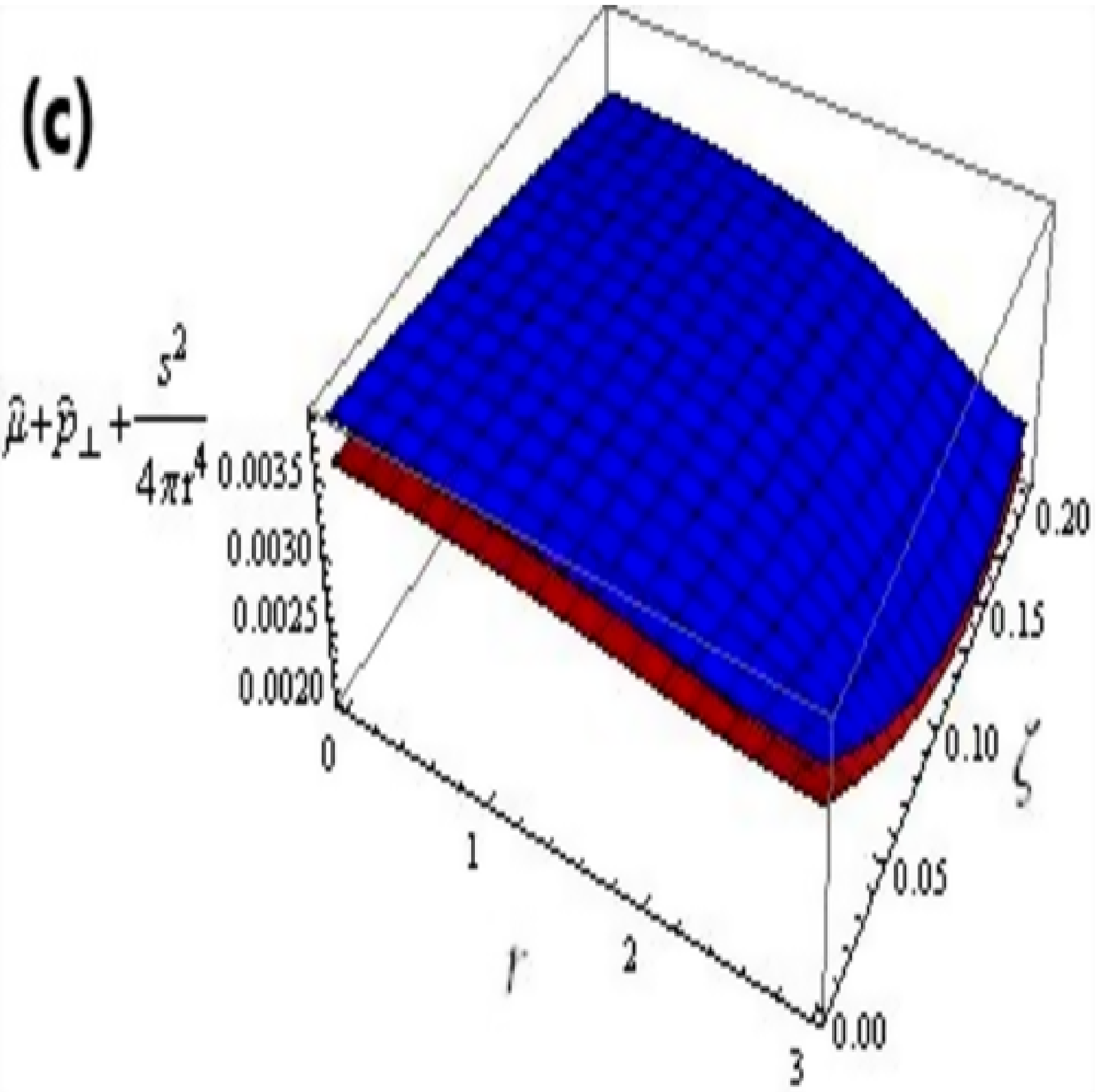,width=0.4\linewidth}\epsfig{file=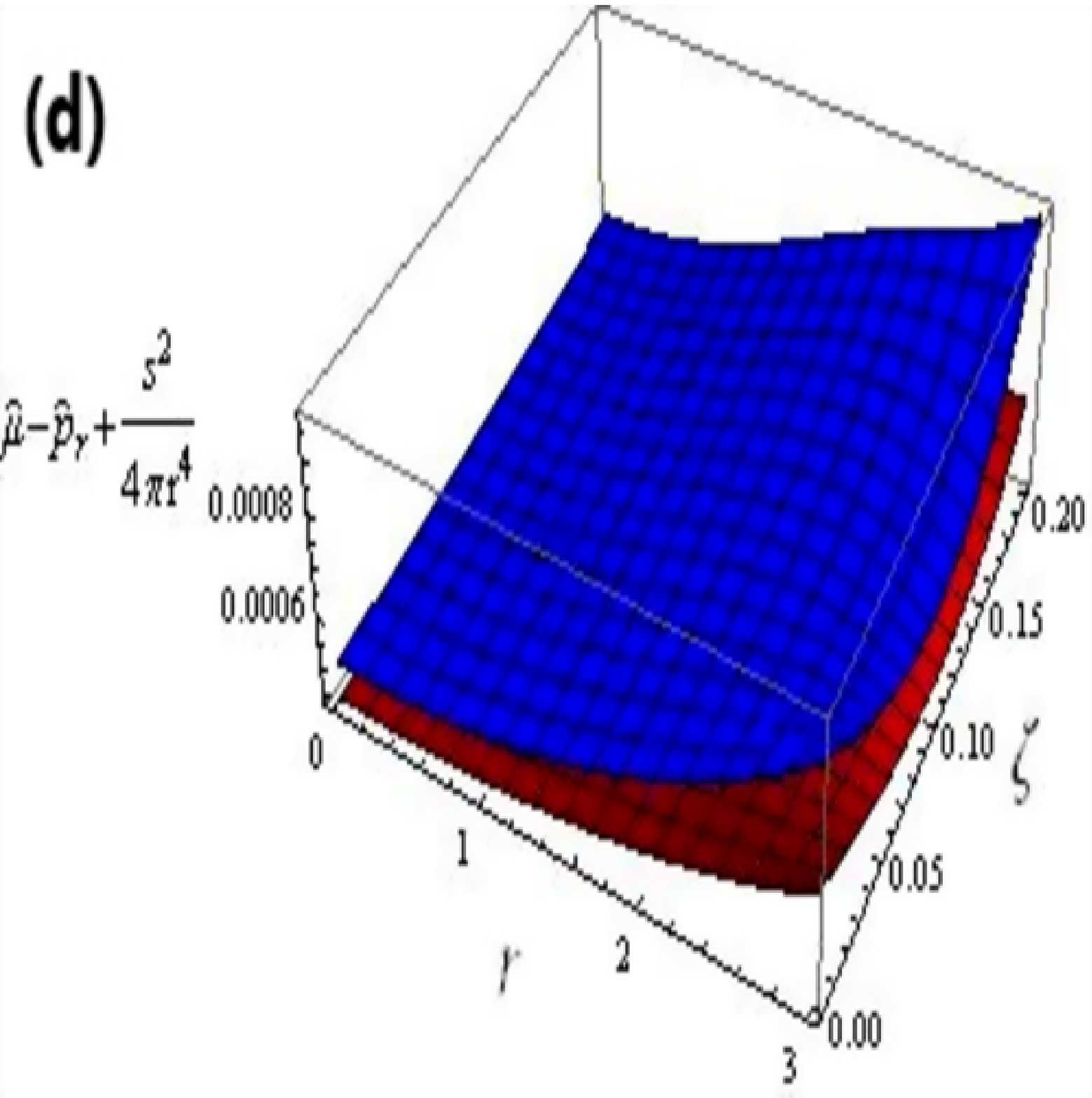,width=0.4\linewidth}
\epsfig{file=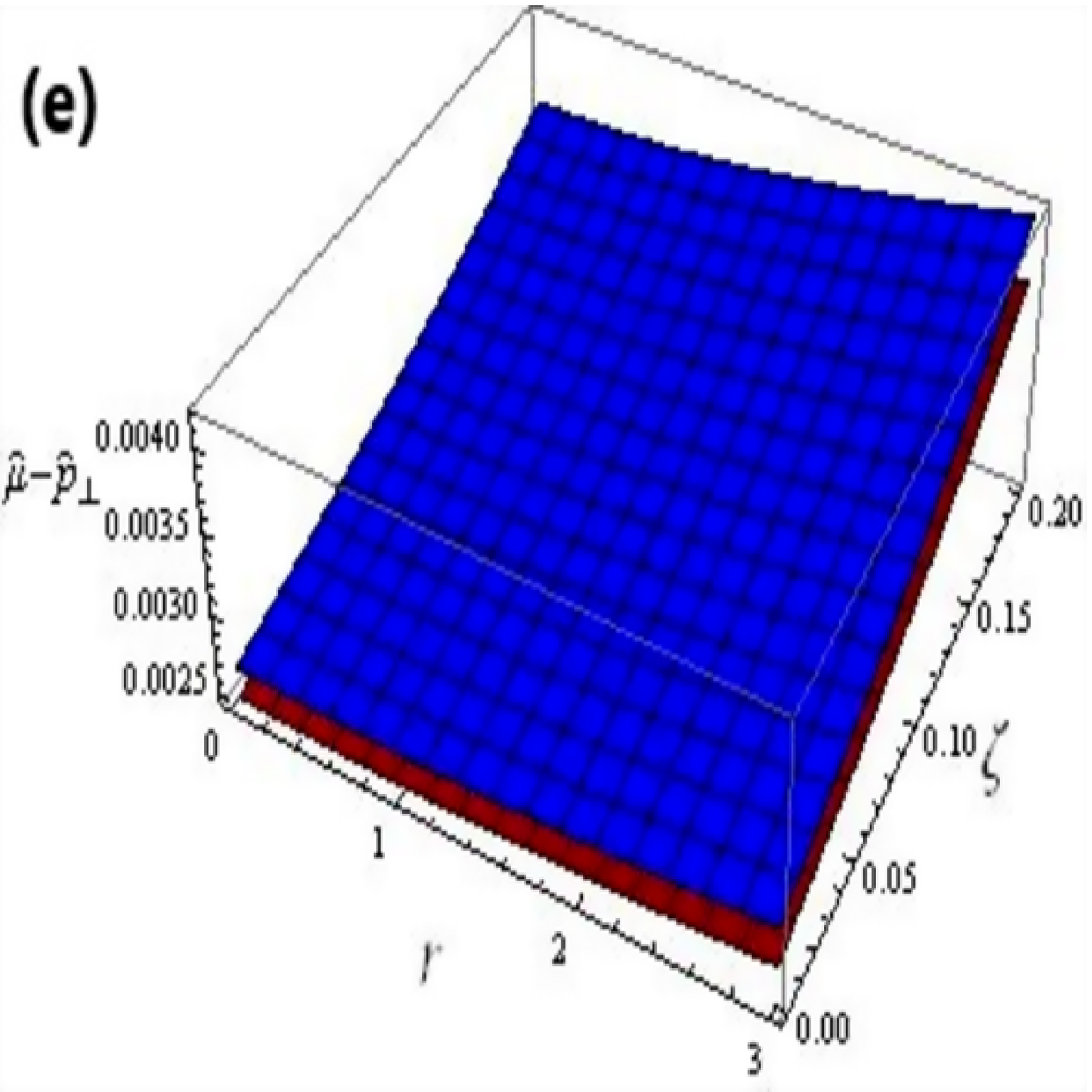,width=0.4\linewidth}\epsfig{file=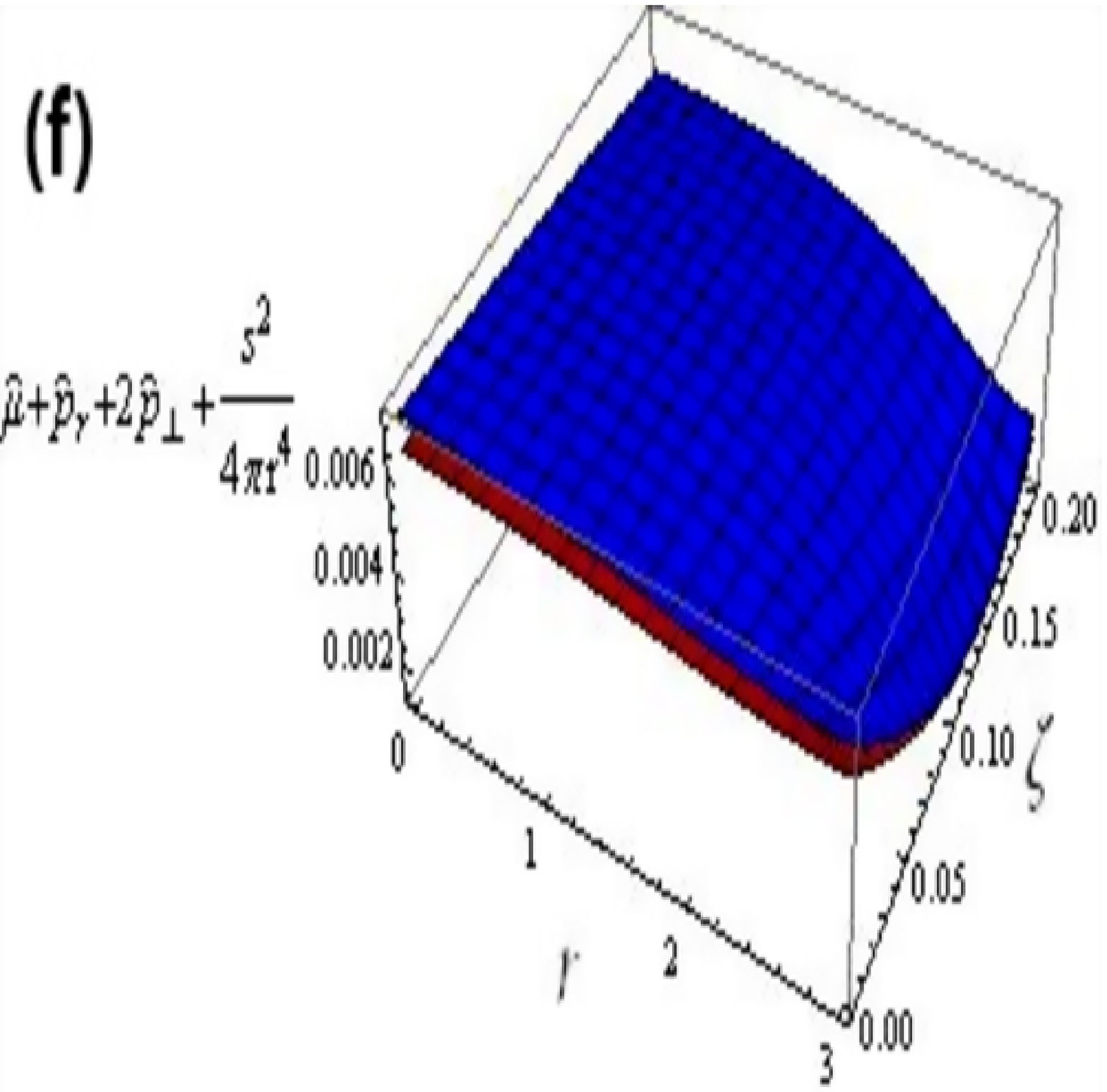,width=0.4\linewidth}
\caption{Plots of energy conditions (in km$^{-2}$) versus $r$ and
$\zeta$ with $\mathcal{S}=0.1$ (Blue), $\mathcal{S}=0.8$ (Red),
$\mathcal{M}=1M_{\bigodot}$ and $H=(0.2)^{-1}M_{\bigodot}$ for
solution-II (\textbf{a}$-$\textbf{f})}
\end{figure}
\begin{figure}\center
\epsfig{file=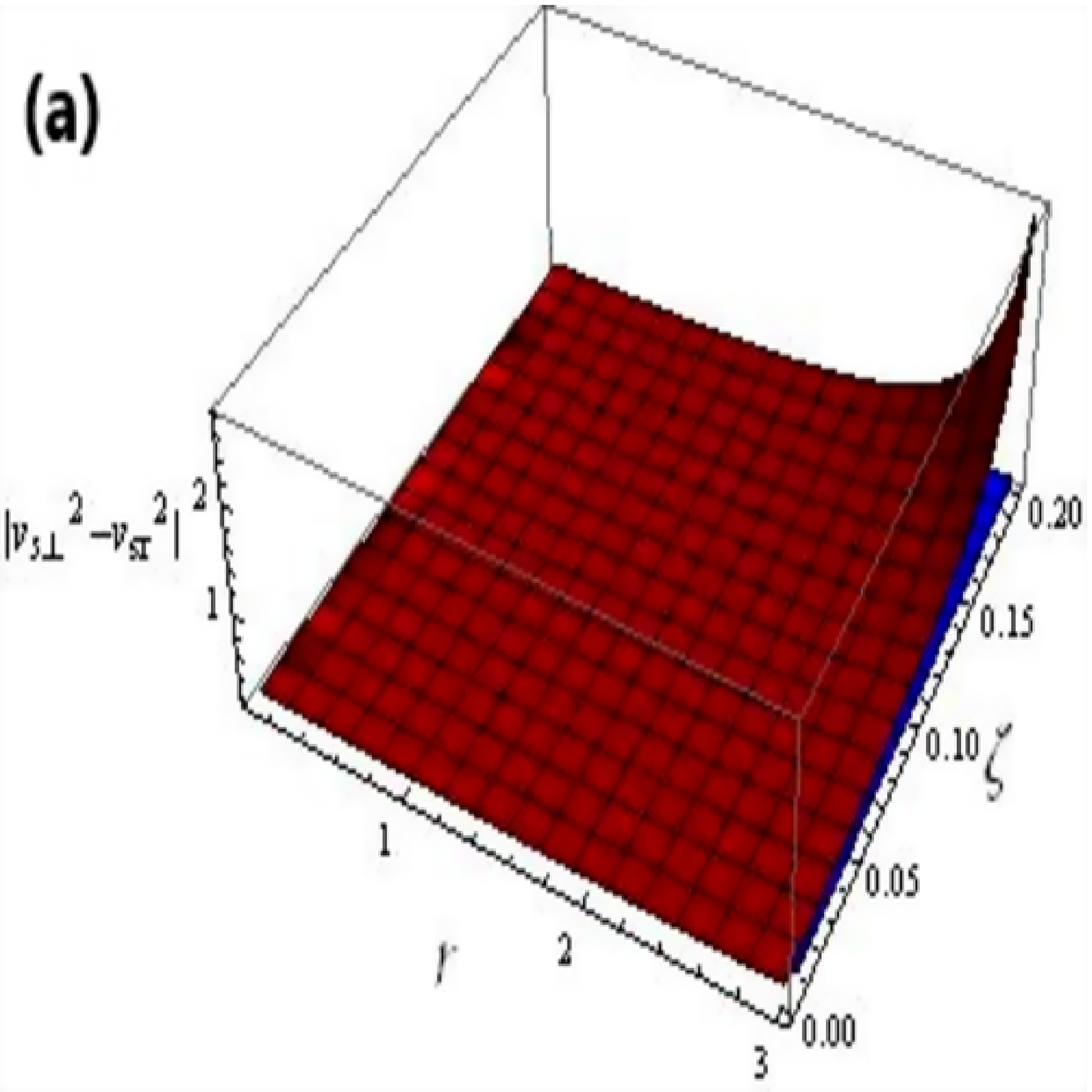,width=0.4\linewidth}\epsfig{file=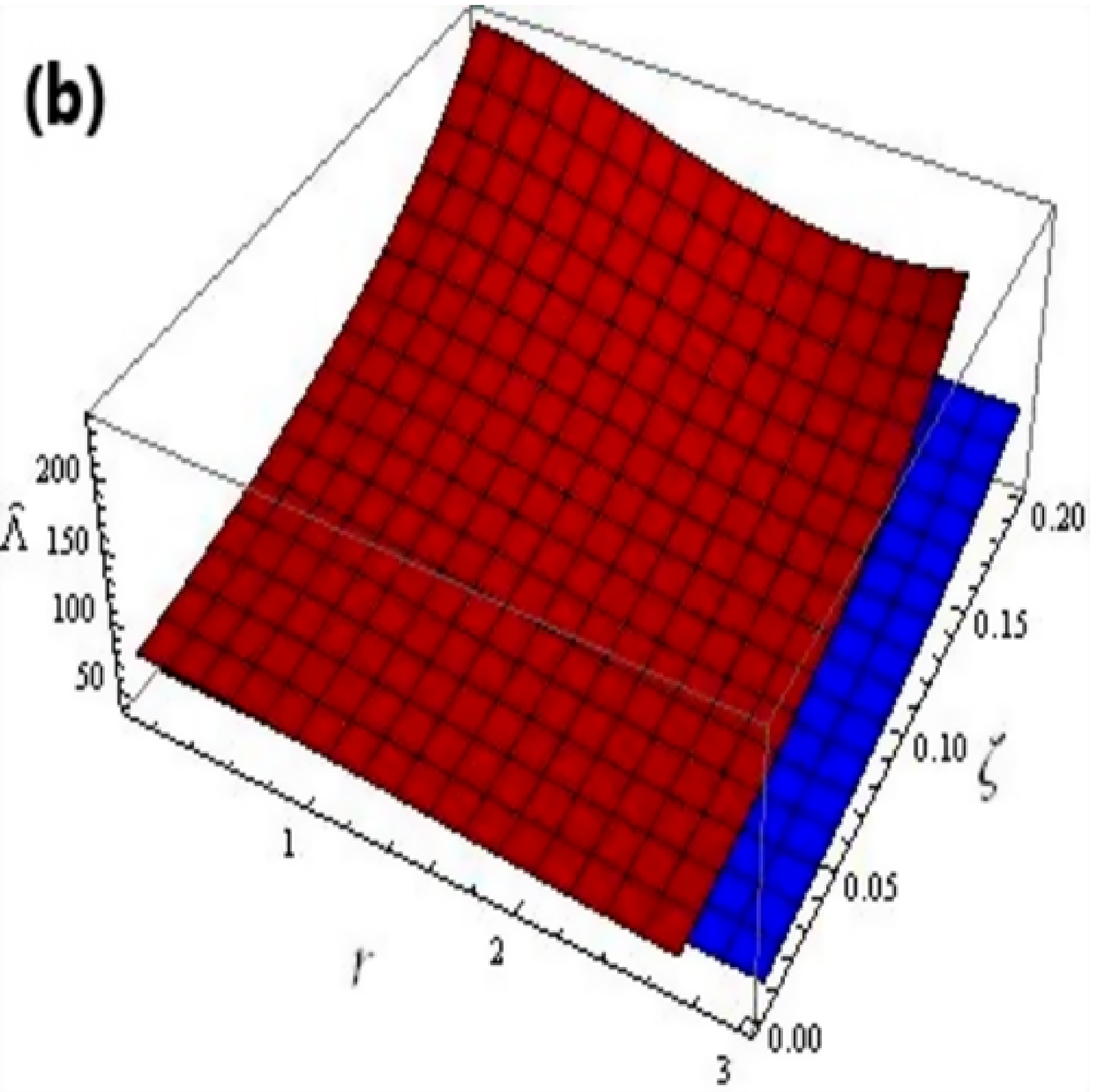,width=0.4\linewidth}
\caption{Plots of $|v_{s\bot}^2-v_{sr}^2|$ (\textbf{a}) and
adiabatic index (\textbf{b}) versus $r$ and $\zeta$ with
$\mathcal{S}=0.1$ (Blue), $\mathcal{S}=0.8$ (Red),
$\mathcal{M}=1M_{\bigodot}$ and $H=(0.2)^{-1}M_{\bigodot}$ for
solution-II}
\end{figure}

\section{Conclusions}

This paper aims to investigate various anisotropic solutions for a
compact spherically symmetric geometry \eqref{g6} with the help of
EGD strategy. For this analysis, we take a linear model
$\mathcal{R}+\varrho \mathcal{Q}$ in
$f(\mathcal{R},\mathcal{T},\mathcal{Q})$ gravitational theory. The
corresponding field equations have been developed and further split
into two sets through the deformation functions. The first set
represents an isotropic configuration, for which we have taken the
isotropic Krori-Barua ansatz in this theory. The unknowns $A,~B$ and
$C$ are computed using the matching conditions. To work out the
second sector \eqref{g21}-\eqref{g23} involving five unknowns, we
have used two constraints to make the system definite. The first one
is the equation of state
$\Upsilon^{0}_{0}=\tau\Upsilon^{1}_{1}+\upsilon\Upsilon^{2}_{2}$,
where $\tau$ and $\upsilon$ are kept fixed, while the other is taken
as pressure-like or density-like, leading to solutions-I and II,
respectively.

To inspect the influence of the decoupling parameter as well as
charge on the obtained solutions, we have discussed the graphical
behavior of effective material variables
$(\hat{\mu},\hat{p}_{r},\hat{p}_{\bot})$, pressure anisotropy
$(\hat{\Pi})$ and energy conditions \eqref{g50} for $\varrho=-0.1$
and $-0.05$. The redshift and compactness factors have also been
found within their respective bounds. The compact geometry
\eqref{g6} becomes more massive with the increment of the decoupling
parameter $\zeta$ for the both solutions, whereas the structure
becomes less dense by increasing charge. We have utilized two
different approaches to analyze the stability of these solutions. It
is found that both solutions provide viable as well as stable
geometry for particular values of $\zeta$ and charge. It is
worthwhile to mention here that solution-I remains stable for the
considered values of charge and $\zeta$, whereas solution-II becomes
less stable with the increment in both these quantities near the
boundary. However, the large values of charge may yield unstable
system analogous to the first solution. We would like to mention
here that this technique provides unstable solution corresponding to
the density-like constraint in GR \cite{40,41} as well as
$f(\mathcal{G})$ theory \cite{41a}. However, our resulting solutions
show physically stable behavior even for larger values of $\zeta$.
Moreover, the anisotropy does not vanish at the center in GR unlike
$f(\mathcal{R},\mathcal{T},\mathcal{Q})$ framework. Thus we conclude
that this modified gravity produces more suitable results. It can be
said that extra force existing in
$f(\mathcal{R},\mathcal{T},\mathcal{Q})$ theory could be the reason
that offers differences of the consequences in this gravity from
those in GR and other modified theories. Finally, for $\varrho=0$,
all our results reduce to GR.

\vspace{0.25cm}

\section*{Appendix A}

The modified matter components appearing in the field equations
\eqref{g8}-\eqref{g10} are given as
\begin{eqnarray}\nonumber
\mathcal{T}_{0}^{0(\mathcal{D})}&=&\frac{1}{8\pi\big(f_{\mathcal{R}}+\mu
f_{\mathcal{Q}}\big)}\left[\mu\left\{f_{\mathcal{Q}}\left(\frac{\chi'^2}{2e^{\beta}}-\frac{\chi'}{re^{\beta}}+\frac{\chi'\beta'}{4e^{\beta}}
-\frac{\chi''}{2e^{\beta}}-\frac{1}{2}\mathcal{R}\right)+f'_{\mathcal{Q}}\left(\frac{\chi'}{2e^{\beta}}\right.\right.\right.\\\nonumber
&-&\left.\left.\frac{\beta'}{4e^{\beta}}+\frac{1}{re^{\beta}}\right)+\frac{f''_{\mathcal{Q}}}{2e^{\beta}}-2f_{\mathcal{T}}\right\}
+\mu'\left\{f_{\mathcal{Q}}\left(\frac{\chi'}{2e^{\beta}}
+\frac{1}{re^{\beta}}-\frac{\beta'}{4e^{\beta}}\right)+\frac{f'_{\mathcal{Q}}}{e^{\beta}}\right\}\\\nonumber
&+&\frac{f_{\mathcal{Q}}\mu''}{2e^{\beta}}+p\left\{f_{\mathcal{Q}}\left(\frac{3\beta'^2}{4e^{\beta}}-\frac{2}{r^2e^{\beta}}-\frac{\beta''}{2e^{\beta}}\right)
-f'_{\mathcal{Q}}\left(\frac{5\beta'}{4e^{\beta}}-\frac{1}{re^{\beta}}\right)+\frac{f''_{\mathcal{Q}}}{2e^{\beta}}\right\}\\\nonumber
&+&p'\left\{f_{\mathcal{Q}}\left(\frac{1}{re^{\beta}}
-\frac{5\beta'}{4e^{\beta}}\right)+\frac{f'_{\mathcal{Q}}}{e^{\beta}}\right\}+\frac{f_{\mathcal{Q}}p''}{2e^{\beta}}+\frac{\mathcal{R}f_{\mathcal{R}}}{2}
+f'_{\mathcal{R}}\left(\frac{\beta'}{2e^{\beta}}-\frac{2}{re^{\beta}}\right)\\\nonumber
&-&\left.\frac{f''_{\mathcal{R}}}{e^{\beta}}-\frac{f}{2}+\frac{q^2}{r^4}\left\{f_{\mathcal{T}}+\frac{f_{\mathcal{Q}}}{4e^{\beta}}\left(\chi'\beta'
-2\chi''-\chi'^2+\frac{4\beta'}{r}\right)\right\}\right],\\\nonumber
\mathcal{T}_{1}^{1(\mathcal{D})}&=&\frac{1}{8\pi\big(f_{\mathcal{R}}+\mu
f_{\mathcal{Q}}\big)}\left[\mu\left(f_{\mathcal{T}}-\frac{f_{\mathcal{Q}}\chi'^2}{4e^{\beta}}
+\frac{f'_{\mathcal{Q}}\chi'}{4e^{\beta}}\right)+\frac{f_{\mathcal{Q}}\mu'\chi'}{4e^{\beta}}+p\left\{f_{\mathcal{T}}
+f_{\mathcal{Q}}\left(\frac{\chi''}{e^{\beta}}\right.\right.\right.\\\nonumber
&-&\left.\left.\frac{\beta'^2}{e^{\beta}}+\frac{\chi'^2}{2e^{\beta}}-\frac{3\chi'\beta'}{4e^{\beta}}-\frac{3\beta'}{re^{\beta}}
+\frac{2}{r^2e^{\beta}}+\frac{1}{2}\mathcal{R}\right)
-f'_{\mathcal{Q}}\left(\frac{\chi'}{4e^{\beta}}+\frac{2}{re^{\beta}}\right)\right\}\\\nonumber
&-&p'f_{\mathcal{Q}}\left(\frac{\chi'}{4e^{\beta}}+\frac{2}{re^{\beta}}\right)+\frac{f}{2}-\frac{\mathcal{R}f_{\mathcal{R}}}{2}
-f'_{\mathcal{R}}\left(\frac{\chi'}{2e^{\beta}}+\frac{2}{re^{\beta}}\right)+\frac{q^2}{r^4}\left\{f_{\mathcal{T}}\right.\\\nonumber
&-&\left.\left.\frac{f_{\mathcal{Q}}}{4e^{\beta}}\left(2\chi''+\chi'^2-\chi'\beta'+\frac{4\chi'}{r}\right)\right\}\right],\\\nonumber
\mathcal{T}_{2}^{2(\mathcal{D})}&=&\frac{1}{8\pi\big(f_{\mathcal{R}}+\mu
f_{\mathcal{Q}}\big)}\left[\mu\left(f_{\mathcal{T}}-\frac{f_{\mathcal{Q}}\chi'^2}{4e^{\beta}}
+\frac{f'_{\mathcal{Q}}\chi'}{4e^{\beta}}\right)+\frac{f_{\mathcal{Q}}\mu'\chi'}{4e^{\beta}}+p\left\{f_{\mathcal{T}}
+f_{\mathcal{Q}}\left(\frac{\beta''}{2e^{\beta}}\right.\right.\right.\\\nonumber
&+&\left.\frac{\chi'}{2re^{\beta}}-\frac{3\beta'^2}{4e^{\beta}}-\frac{\beta'}{2re^{\beta}}+\frac{1}{r^2e^{\beta}}-\frac{2}{r^2}
+\frac{1}{2}\mathcal{R}\right)+f'_{\mathcal{Q}}\left(\frac{3\beta'}{2e^{\beta}}-\frac{\chi'}{4e^{\beta}}-\frac{3}{re^{\beta}}\right)\\\nonumber
&-&\left.\frac{f''_{\mathcal{Q}}}{e^{\beta}}\right\}+p'\left\{f_{\mathcal{Q}}\left(\frac{3\beta'}{2e^{\beta}}-\frac{\chi'}{4e^{\beta}}
-\frac{3}{re^{\beta}}\right)-\frac{2f'_{\mathcal{Q}}}{e^{\beta}}\right\}-\frac{f_{\mathcal{Q}}p''}{e^{\beta}}
-\frac{\mathcal{R}f_{\mathcal{R}}}{2}+\frac{f}{2}\\\nonumber
&+&\left.f'_{\mathcal{R}}\left(\frac{\beta'}{2e^{\beta}}-\frac{\chi'}{2e^{\beta}}
-\frac{1}{re^{\beta}}\right)-\frac{f''_{\mathcal{R}}}{e^{\beta}}\right].
\end{eqnarray}
The term $\Omega$ in Eq.\eqref{g12} which occurs due to modified
gravity is
\begin{align}
\nonumber \Omega &=
\frac{2}{\left(\mathcal{R}f_{\mathcal{Q}}+2(8\pi+f_{\mathcal{T}})\right)}\left[f'_{\mathcal{Q}}e^{-\beta}\left(p-\frac{q^2}{8\pi
r^4}\right)\left(\frac{1}{r^2}-\frac{e^\beta}{r^2}
+\frac{\chi'}{r}\right)+f_{\mathcal{Q}}e^{-\beta}\right.\\\nonumber
&\times\left(p-\frac{q^2}{8\pi
r^4}\right)\left(\frac{\chi''}{r}-\frac{\chi'}{r^2}-\frac{\beta'}{r^2}-\frac{\chi'\beta'}{r}-\frac{2}{r^3}+\frac{2e^\beta}{r^3}\right)+\left(p'-\frac{qq'}{4\pi
r^4}+\frac{q^2}{2\pi r^5}\right)\\\nonumber
&\times\left\{f_{\mathcal{Q}}e^{-\beta}\left(\frac{\chi'\beta'}{8}
-\frac{\chi''}{8}-\frac{\chi'^2}{8}+\frac{\beta'}{2r}+\frac{\chi'}{2r}+\frac{1}{r^2}-\frac{e^{\beta}}{r^2}\right)+\frac{3}{4}f_{\mathcal{T}}\right\}
+\left(p-\frac{q^2}{8\pi r^4}\right)\\\nonumber &\times
f'_{\mathcal{T}}-\left(\mu+\frac{q^2}{8\pi r^4}\right)
f'_{\mathcal{T}}-\left(\mu'+\frac{qq'}{4\pi r^4}-\frac{q^2}{2\pi
r^5}\right)\left\{\frac{f_{\mathcal{Q}}e^{-\beta}}{8}\left(\chi'^2-\chi'\beta'+2\chi''\right.\right.\\\nonumber
&+\left.\left.\frac{4\chi'}{r}\right)
+\frac{3f_{\mathcal{T}}}{2}\right\}-\left(\frac{e^{-\beta}}{r^2}-\frac{1}{r^2}+\frac{\chi'e^{-\beta}}{r}\right)\left\{\left(\mu'+\frac{qq'}{4\pi
r^4}-\frac{q^2}{2\pi
r^5}\right)f_{\mathcal{Q}}+f'_{\mathcal{Q}}\right.\\\nonumber
&\times\left.\left.\left(\mu+\frac{q^2}{8\pi
r^4}\right)\right\}-\frac{1}{2}\left(f'_{\mathcal{Q}}\mathcal{R}+f_{\mathcal{Q}}\mathcal{R}'+2f'_{\mathcal{T}}\right)
\left(p-\frac{q^2}{8\pi r^4}\right)\right].
\end{align}


\vspace{0.5cm}

\begin{thebibliography}{43}

\bibitem{2} S Nojiri and S D Odintsov \emph{Phys. Rev.
D} \textbf{68} 123512 (2003)

\bibitem{2a} G Cognola, E Elizalde, S Nojiri, S D Odintsov and S Zerbini \emph{J. Cosmol. Astropart.
Phys.} \textbf{2005} 010 (2005)

\bibitem{2b} Y S Song, W Hu and I Sawicki \emph{Phys.
Rev. D} \textbf{75} 044004 (2007)

\bibitem{2c} M Akbar and R G Cai \emph{Phys.
Lett. B} \textbf{648} 243 (2007)

\bibitem{8} S Capozziello, M De Laurentis, S D Odintsov and A Stabile \emph{Phys. Rev. D} \textbf{83} 064004 (2011)

\bibitem{9} M Sharif and H R Kausar \emph{J. Cosmol. Astropart.
Phys.} \textbf{2011} 022 (2011)

\bibitem{9a} M Sharif and H R Kausar \emph{J. Phys. Soc. Japan} \textbf{80} 044004 (2011)

\bibitem{9b} S Arapo{\u{g}}lu, C Deliduman and K Y Ek{\c{s}}i \emph{J. Cosmol.
Astropart. Phys.} \textbf{2011} 020 (2011)

\bibitem{9c} R Goswami, A M Nzioki, S D Maharaj and S G Ghosh \emph{Phys. Rev. D} \textbf{90} 084011 (2014)

\bibitem{9d} M Sharif and Z Yousaf \emph{Astropart. Phys.} \textbf{56} 19 (2014)

\bibitem{9e} M Sharif and Z Yousaf \emph{Astrophys. Space Sci.} \textbf{354} 471 (2014)

\bibitem{9f} A V Astashenok, S Capozziello and S D Odintsov \emph{Phys. Rev. D} \textbf{89} 103509 (2014)

\bibitem{9g} A V Astashenok, S Capozziello and S D Odintsov \emph{J. Cosmol. Astropart. Phys.} \textbf{2015} 001 (2015)

\bibitem{10} O Bertolami, C G Boehmer, T Harko and F S N Lobo \emph{Phys. Rev. D} \textbf{75} 104016 (2007)

\bibitem{20} T Harko, F S N Lobo, S Nojiri and S D Odintsov \emph{Phys. Rev. D} \textbf{84} 024020 (2011)

\bibitem{22} Z Haghani, T Harko, F S N Lobo, H R Sepangi and S Shahidi \emph{Phys. Rev. D} \textbf{88} 044023 (2013)

\bibitem{22a} M Sharif and M Zubair \emph{J. Cosmol. Astropart. Phys.} \textbf{2013} 042 (2013)

\bibitem{22b} M Sharif and M Zubair \emph{J. High Energy Phys.} \textbf{2013} 79 (2013)

\bibitem{23} S D Odintsov and D S{\'a}ez-G{\'o}mez \emph{Phys. Lett. B} \textbf{725} 437 (2013)

\bibitem{24} I Ayuso, J B Jim{\'e}nez and A De la Cruz-Dombriz \emph{Phys. Rev. D} \textbf{91} 104003 (2015)

\bibitem{25} E H Baffou, M J S Houndjo and J Tosssa \emph{Astrophys. Space Sci.} \textbf{361} 376 (2016)

\bibitem{26} Z Yousaf, M Z Bhatti and T Naseer \emph{Eur. Phys. J. Plus} \textbf{135} 353 (2020)

\bibitem{26a} Z Yousaf, M Z Bhatti and T Naseer \emph{Phys. Dark Universe} \textbf{28} 100535 (2020)

\bibitem{26b} Z Yousaf, M Z Bhatti and T Naseer \emph{Int. J. Mod. Phys. D} \textbf{29} 2050061 (2020)

\bibitem{26c} Z Yousaf, M Z Bhatti and T Naseer \emph{Ann. Phys.} \textbf{420} 168267 (2020)

\bibitem{26d} Z Yousaf, M Z Bhatti, T Naseer and I Ahmad \emph{Phys. Dark Universe} \textbf{29} 100581 (2020)

\bibitem{26e} Z Yousaf, M Y Khlopov, M Z Bhatti and T Naseer \emph{Mon. Not. R. Astron. Soc.} \textbf{495} 4334 (2020)

\bibitem{27} B Das, P C Ray, I Radinschi, F Rahaman and S Ray \emph{Int. J. Mod. Phys. D} \textbf{20} 1675 (2011)

\bibitem{27a} J M Sunzu, S D Maharaj and S Ray \emph{Astrophys. Space Sci.} \textbf{352} 719 (2014)

\bibitem{28} M H Murad \emph{Astrophys. Space Sci.} \textbf{361} 20 (2016)

\bibitem{28a} N Pant, R N Mehta and M J Pant \emph{Astrophys. Space Sci.} \textbf{332} 473 (2011)

\bibitem{28b} Y K Gupta and S K Maurya \emph{Astrophys. Space Sci.} \textbf{332} 155 (2011)

\bibitem{28c} M Sharif and M Z Bhatti \emph{Astrophys. Space Sci.} \textbf{347} 337 (2013)

\bibitem{28d} M Sharif and Z Yousaf \emph{Phys. Rev. D} \textbf{88} 024020 (2013)

\bibitem{28e} K N Singh and N Pant \emph{Astrophys. Space Sci.} \textbf{358} 1 (2015)

\bibitem{28f} M Sharif and S Sadiq \emph{Eur. Phys. J. C} \textbf{76} 1 (2016)

\bibitem{28g} M Sharif and A Siddiqa \emph{Eur. Phys. J. Plus} \textbf{132} 1 (2017)

\bibitem{28h} M Sharif and A Waseem \emph{Gen. Relativ. Gravit.} \textbf{50} 1 (2018)

\bibitem{28i} M Sharif and A Waseem \emph{Eur. Phys. J. Plus} \textbf{131} 1 (2016)

\bibitem{28j} M Sharif and A Waseem \emph{Can. J. Phys.} \textbf{94} 1024 (2016)

\bibitem{28k} S K Maurya, A Errehymy, D Deb, F Tello-Ortiz and M Daoud \emph{Phys. Rev. D} \textbf{100} 044014 (2019)

\bibitem{28l} M F Shamir and I Fayyaz \emph{Theor. Math. Phys.} \textbf{202} 112 (2020)

\bibitem{29} J Ovalle \emph{Mod. Phys. Lett. A} \textbf{23} 3247 (2008)

\bibitem{30} J Ovalle and F Linares \emph{Phys. Rev. D} \textbf{88} 104026 (2013)

\bibitem{31} R Casadio, J Ovalle and R Da Rocha \emph{Class. Quantum Grav.} \textbf{32} 215020 (2015)

\bibitem{32} J Ovalle \emph{Phys. Rev. D} \textbf{95} 104019 (2017)

\bibitem{33} J Ovalle, R Casadio, R da Rocha, A Sotomayor and Z. Stuchl{\'\i}k \emph{Eur. Phys. J. C} \textbf{78} 1 (2018)

\bibitem{34} L Gabbanelli, {\'A} Rinc{\'o}n and C Rubio \emph{Eur. Phys. J. C} \textbf{78} 370 (2018)

\bibitem{34a} M Estrada and F Tello-Ortiz \emph{Eur. Phys. J. Plus} \textbf{133} 1 (2018)

\bibitem{35} M Sharif and S Sadiq \emph{Eur. Phys. J. C} \textbf{78} 410 (2018)

\bibitem{36} M Sharif and S Saba \emph{Eur. Phys. J. C} \textbf{78} 921 (2018)

\bibitem{36a} M Sharif and S Saba \emph{Chin. J. Phys.} \textbf{59} 481 (2019)

\bibitem{36b} M Sharif and A Waseem \emph{Ann. Phys.} \textbf{405} 14 (2019)

\bibitem{36c} M Sharif and A Waseem \emph{Chin. J. Phys.} \textbf{60} 426 (2019)

\bibitem{37} K N Singh, S K Maurya, M K Jasim and F Rahaman \emph{Eur. Phys. J. C} \textbf{79} 1 (2019)

\bibitem{37a} S Hensh and Z Stuchl{\'\i}k \emph{Eur. Phys. J. C} \textbf{79} 1 (2019)

\bibitem{38} J Ovalle \emph{Phys. Lett. B} \textbf{788} 213 (2019)

\bibitem{39} E Contreras and P Bargue{\~n}o \emph{Class. Quantum Gravity} \textbf{36} 215009 (2019)

\bibitem{40} M Sharif and Q Ama-Tul-Mughani \emph{Ann. Phys.} \textbf{415} 168122 (2020)

\bibitem{41} M Sharif and Q Ama-Tul-Mughani \emph{Chin. J. Phys.} \textbf{65} 207 (2020)

\bibitem{41a} M Sharif and S Saba \emph{Int. J. Mod. Phys. D} \textbf{29} 2050041 (2020)

\bibitem{41b} M Sharif and T Naseer \emph{Chin. J. Phys.} \textbf{73} 179 (2021)

\bibitem{41c} T Naseer and M Sharif \emph{Universe} \textbf{8} 62 (2022)

\bibitem{41d} M Sharif and A Majid \emph{Phys. Dark Universe} \textbf{32} 100803 (2021)

\bibitem{41e} M Sharif and A Majid \emph{Phys. Scr.} \textbf{96} 035002 (2021)

\bibitem{41f} M Sharif and A Majid \emph{Phys. Scr.} \textbf{96} 045003 (2021)

\bibitem{42} K D Krori and J Barua \emph{J. Phys. A: Math. Gen.} \textbf{8} 508 (1975)

\bibitem{42a} H A Buchdahl \emph{Phys. Rev.} \textbf{116} 1027 (1959)

\bibitem{42b} B V Ivanov \emph{Phys. Rev. D} \textbf{65} 104011 (2002)

\bibitem{42c} H Abreu, H Hernandez and L A Nunez \emph{Class. Quantum Gravit.} \textbf{24} 4631 (2007)

\bibitem{42ca} L Herrera \emph{Phys. Lett. A} \textbf{165} 206 (1992)

\bibitem{42d} H Heintzmann and W Hillebrandt \emph{Astron. Astrophys.} \textbf{38} 51 (1975)

\bibitem{42e} W Hillebrandt and K O Steinmetz \emph{Astron. Astrophys.} \textbf{53} 283 (1976)

\bibitem{42f} I Bombaci \emph{Astron. Astrophys.} \textbf{305} 871 (1996)
\end{thebibliography}
\end{document}